\documentclass[pre,aps,floats,floatfix,superscriptaddress,nofootinbib,showpacs]{revtex4}
\usepackage[english]{babel}
\usepackage{amsmath,amssymb,amsfonts}  
\usepackage[dvips]{graphicx}
\usepackage{amsmath}
\usepackage{fancyhdr}

\usepackage{bm}
\newcommand{\be}{\begin{equation}}
\newcommand{\ee}{\end{equation}}

\begin{document}

\title{Brownian motion under annihilation dynamics}
\author{Mar\'ia Isabel Garc\'ia de Soria}
\affiliation{Universit\'e Paris-Sud, LPTMS, UMR 8626, Orsay Cedex, F-91405 and
CNRS, Orsay, F-91405}
\author{Pablo Maynar}
\affiliation{Laboratoire de Physique Th\'eorique (CNRS
  UMR 8627), B\^atiment 210, Universit\'e Paris-Sud, 91405 Orsay cedex,
  France}
\affiliation{F\'{\i}sica Te\'{o}rica, Universidad de Sevilla,
Apartado de Correos 1065, E-41080, Sevilla, Spain}
\author{Emmanuel Trizac}
\affiliation{Universit\'e Paris-Sud, LPTMS, UMR 8626, Orsay Cedex, F-91405 and
CNRS, Orsay, F-91405}

\date{\today }

\begin{abstract}
The behavior of a heavy tagged intruder immersed in a bath of particles 
evolving under ballistic annihilation dynamics is investigated. The 
Fokker-Planck equation for this system is derived and the 
peculiarities of the corresponding diffusive behavior are worked out. 
In the long time limit, the intruder velocity distribution function approaches a
Gaussian form, but with a different temperature from its bath
counterpart. As a consequence of the continuous decay of particles in the bath, 
the mean squared displacement increases exponentially in the collision per
particle time scale. Analytical results are finally successfully tested
against Monte Carlo numerical simulations.

\end{abstract}

\pacs{51.10.+y,05.20.Dd,82.20.Nk}
\maketitle

\begin{section}{Introduction}
In recent
years, there has been some interest for systems where particles annihilate
ballistically \cite{bnklr,bek,ks,t02,ptd02,cdt04,cdt04s,llf06}. 
In these
studies, the model considered consists of an ensemble of hard particles which evolve
freely until a binary encounter, which leads either to the annihilation of the
colliding partners with probability $p$,  or to an elastic collision 
with probability $1-p$. 
For this probabilistic annihilation model, there are no collisional invariants, 
and numerical simulations have shown that for a broad class of initial
conditions, the system reaches an homogeneous state in which all the time
dependence of the one particle distribution function is encoded in the density
and temperature (defined as the second velocity moment of the distribution
function) \cite{t02,ptd02}. This is the so-called ``Homogeneous Decay State''. 
Such a behavior resembles the one of granular fluids (see \cite{bte05} and 
references therein) where, if
the system is stable, it evolves into an homogeneous cooling state, in which all
the time dependence is borne by the granular temperature (in this case the
density is conserved) \cite{gs95}. For the annihilation model, the 
hydrodynamic equations have been derived using the Chapmann-Enskog method
\cite{chapman60} by the usual assumption of the existence of a 
``normal solution'',  whose space
and time dependence occurs only through the hydrodynamic fields
\cite{cdt04}. Recently, the hydrodynamic equations linearized around the
homogeneous decay state have been derived relaxing such an assumption \cite{gmsbt08}. 
Nevertheless, it must be assumed that there is scale
separation, i.e that the spectrum of 
the linearized
Boltzmann collision operator is such that the eigenvalues associated to the
hydrodynamic excitations are separated from the faster ``kinetic
eigenvalues''. Although this property is valid for elastic collisions 
\cite{mclennan}, it has not been proven for the probabilistic ballistic annihilation
model in general, but only
for Maxwell molecules \cite{ernst} and for $p$ smaller than a given threshold \cite{gmsbt08}.

The objective in this paper is to study the simplest transport process in this
system in which we can rigorously prove that there is scale
separation. We will consider a tagged particle in a fluid in the
homogeneous decay state, but collisions between the tagged particle and the particles of the fluid
will be always elastic. 
The equation for the tagged particle is the Boltzmann-Lorentz equation 
\cite{resibois,mclennan} which depends on the one particle distribution
function of the bath. In the limit of asymptotically large relative
mass for the tagged particle, this equation reduces to a Fokker-Planck
equation which depends on the time-dependent density and temperature of 
the bath. Due to the structure of this equation, we can prove that there is 
scale separation and that, in the long-time limit, the velocity distribution
function of the tagged particle approaches a Gaussian distribution but with a
 temperature that differs from that of the bath. 
A similar breakdown of equipartition has been reported for a 
heavy particle in a granular bath 
\cite{mp99,bds99,brgd99,dbl02th,bt02}, a problem that can be mapped 
onto an elastic situation \cite{sd06}, at variance with the situation
under scrutiny here. We also study the diffusion
of the heavy particle and identify the diffusion coefficient as a Green-Kubo
formula in terms of the velocity autocorrelation function. 
Finally, we perform Monte Carlo numerical simulations in
order to test our theoretical results. 

\end{section}

\begin{section}{Fokker-Planck equation}

We consider a tagged particle of mass $m$ and diameter $\sigma$ immersed
in a low-density gas. This gas is composed of hard spheres or disks of mass
$m_g$ and diameter $\sigma_g$ which move ballistically until one particle
meets another one; such binary encounters lead to the annihilation of the
colliding partners with probability $p$ or to an elastic collision with
probability $1-p$ \cite{bnklr,bek,ks,t02,ptd02,llf06}. 
Collisions between the particles of the gas and the
tagged particle are always elastic. 

\subsection{From Boltzmann-Lorentz to Fokker-Planck}
The evolution equation for the probability density 
$F(\mathbf{r},\mathbf{v},t)$ of the tagged particle is the 
Boltzmann-Lorentz equation \cite{resibois,mclennan}
\begin{equation}\label{Boltzmann-Lorentz}
\left(\frac{\partial}{\partial t}+\mathbf{v}\cdot\nabla\right)
F(\mathbf{r},\mathbf{v},t)=J[\mathbf{r},\mathbf{v},t\vert F,f],
\end{equation}
where the collision operator is given by
\begin{eqnarray}\label{op_coll}
J[\mathbf{r},\mathbf{v},t\vert F,f]&=&\sigma_0^{d-1}\int\!\!d\mathbf{v}_1
\!\!\int\!\!d\boldsymbol{\hat{\sigma}}
\Theta(\mathbf{g}\cdot\boldsymbol{\hat{\sigma}})
(\mathbf{g}\cdot\boldsymbol{\hat{\sigma}})
\left\{ F(\mathbf{r},\mathbf{v}^*,t)f(\mathbf{r},\mathbf{v}^*_1,t)\right.
\nonumber\\
&-&\left.F(\mathbf{r},\mathbf{v},t)f(\mathbf{r},\mathbf{v}_1,t)\right\}.
\end{eqnarray}
Here $f(\mathbf{r},\mathbf{v},t)$ is the distribution function of the 
particles 
in the gas, $d$ is the space dimension, $\mathbf{g}=\mathbf{v}-\mathbf{v}_1$
is the relative velocity, 
$\Theta$ is the Heaviside 
step function, $\boldsymbol{\hat{\sigma}}$ is a unit vector pointing from the 
centre of the gas particle to the centre of the tagged particle at
contact, and $\sigma_0=\frac{\sigma+\sigma_g}{2}$. The precollisional 
velocities $\mathbf{v}^*$ and $\mathbf{v}_1^*$ are given by
\begin{eqnarray}
\mathbf{v}^{*}=\mathbf{v}-\frac{2\Delta}{1+\Delta}
(\mathbf{g}\cdot\boldsymbol{\hat{\sigma}})\boldsymbol{\hat{\sigma}},\\
\mathbf{v}_1^{*}=\mathbf{v}_1+\frac{2}{1+\Delta}
(\mathbf{g}\cdot\boldsymbol{\hat{\sigma}})\boldsymbol{\hat{\sigma}},
\end{eqnarray}
with $\Delta=m_g/m$ the (gas/tagged particle) mass ratio.

We shall consider that the gas is in the homogeneous decay state, so its distribution 
function has the scaling form \cite{ptd02}
\begin{equation}\label{scalingHDS}
f_H(\mathbf{v}_1,t)=\frac{n_g(t)}{v_g^d(t)}\chi_H({c}_1), \qquad
\mathbf{c}_1=\frac{\mathbf{v}_1}{v_g(t)},
\end{equation}
where $n_g(t)$ is the number density of the gas, 
$v_g(t)=\left(\frac{2T_g(t)}{m_g}\right)^{1/2}$ is the thermal velocity of the 
particles in the gas and $\chi_H$ is an isotropic function depending only on
the modulus $c=|\mathbf{c}|$ of the rescaled velocity. It can be seen that the
homogeneous density and temperature obey the following equations \cite{cdt04}
\begin{eqnarray}\label{ec_nH}
\frac{\partial n_g(t)}{\partial t}&=&-p\nu_g(t)\zeta_nn_g(t),\\
\label{ec_TH}
\frac{\partial T_g(t)}{\partial t}&=&-p\nu_g(t)\zeta_TT_g(t),
\end{eqnarray} 
where we have introduced the collision frequency 
of the corresponding hard sphere fluid in equilibrium
(with same temperature and density)
\begin{equation}
\nu_g(t)=\frac{n_g(t) \,T^{1/2}_g(t) \,\sigma_g^{d-1}}{ m^{1/2}}
\frac{8\pi^{\frac{d-1}{2}}}{(d+2)\Gamma(d/2)}.
\label{ec_nuH}
\end{equation}
Here the dimensionless decay rates $\zeta_n$ and $\zeta_T$ are functionals of
the distribution function and  are approximately known in the first Sonine
approximation \cite{t02,cdt04s}, see Appendix \ref{app:A}. Equations
(\ref{ec_nH}) and (\ref{ec_TH}) can be integrated to obtain the following
power laws for the decay of the density and temperature
\begin{eqnarray}\label{nH_sol}
n_g(t)&=&n_g(0)\left[1+\nu_g(0)p(\zeta_n+\zeta_T/2)t\right]
^{-\frac{2\zeta_n}{2\zeta_n+\zeta_T}},\\
T_g(t)&=&T_g(0)\left[1+\nu_g(0)p(\zeta_n+\zeta_T/2)t\right]
^{-\frac{2\zeta_T}{2\zeta_n+\zeta_T}}.\label{TH_sol}
\end{eqnarray}
As a consequence, we get $n_g T_g^{1/2} \propto t^{-1}$, a simplified form of a scaling
relation common to all ballistically controlled processes
\cite{TH95}.

We next study the evolution equation for the tagged particle in
the limit of large relative mass for the tagged particle. In 
the limit $\Delta \ll 1$, it is possible to expand the collision operator 
$J[\mathbf{r},\mathbf{v},t\vert F,f]$ in powers of $\Delta$. In Appendix 
\ref{appendixA} it
is shown that the leading order is 
\begin{equation}\label{ap_c_o}
J[\mathbf{r},\mathbf{v},t\vert F,f]\simeq
\frac{\partial}{\partial\mathbf{v}}\cdot
[\mathbf{A}(\mathbf{v})F(\mathbf{r},\mathbf{v},t)]+\frac{1}{2}
\frac{\partial}{\partial\mathbf{v}}\frac{\partial}{\partial\mathbf{v}}:
[N(\mathbf{v})F(\mathbf{r},\mathbf{v},t)], 
\end{equation}
where 
\begin{equation}
\mathbf{A}(\mathbf{v},t)=\gamma(t)\mathbf{v}, \qquad
N(\mathbf{v},t)=2\bar{\gamma}(t)I,
\end{equation}
and $I$ is the second order unit tensor. The definitions of $\gamma$ and
$\bar{\gamma}$ are respectively
\begin{eqnarray}
\gamma(t)&=&\gamma_e[n_g(t),T_g(t)]a(p),\\
\bar{\gamma}(t)&=&\gamma_e[n_g(t),T_g(t)]a(p)b(p)\frac{T_g(t)}{m},
\end{eqnarray}
The friction coefficient $\gamma_e(t)$ is the same as for elastic bodies, 
and appears here as a function of the time-dependent density $n_g(t)$ and temperature 
$T_g(t)$
\begin{equation}
\gamma_e[n_g(t),T_g(t)]=\frac{4\pi^{\frac{d-1}{2}}}{d\Gamma(d/2)}\Delta^{1/2}
n_g(t)\left(\frac{2T_g(t)}{m}\right)^{1/2}\sigma_0^{d-1},
\end{equation}
with $a(p)$ and $b(p)$ functionals of the distribution function of the bath 
which depend only on the parameter $p$
\begin{eqnarray}\label{ec.a}
a(p)&=&\frac{\Gamma(d/2)}{\Gamma((d+1)/2)}\int\!\!d\mathbf{c}_1
\chi_H(c_1)c_1,\\
\label{ec.b}
b(p)&=&\frac{2}{d+1}\frac{\int\!\!d\mathbf{c}_1\chi_H(c_1)c_1^3}
{\int\!\!d\mathbf{c}_1\chi_H(c_1)c_1}.
\end{eqnarray}
In Appendix \ref{appendixA}, it is shown that the two terms on the right hand side of 
equation (\ref{ap_c_o}) are both of 
order $n_gv_g\sigma_0^{d-1}\Delta$, while the other contributions in the
Kramers-Moyal expansion are at least of order 
$n_gv_g\sigma_0^{d-1}\Delta^{3/2}$. In the same Appendix, the expressions for $a(p)$ and $b(p)$ are evaluated to first order in a Sonine expansion. 

Taking into account the approximate expression for the collision operator, Eq.
(\ref{ap_c_o}), it is possible to write the Boltzmann-Lorentz equation as a 
Fokker-Planck equation for asymptotically small $\Delta$ 
\begin{equation}\label{Fokker-Planck}
\left[\frac{\partial}{\partial t}+\mathbf{v}\cdot\nabla\right]
F(\mathbf{r},\mathbf{v},t)=\gamma_e(t)a(p)
\frac{\partial}{\partial\mathbf{v}}\cdot
\left[\mathbf{v}+b(p)\frac{T_g(t)}{m}
\frac{\partial}{\partial\mathbf{v}}\right]F(\mathbf{r},\mathbf{v},t).
\end{equation}
As in the inelastic case, the Einstein relation is violated due to the fact
that the distribution function of the bath is not Maxwellian
\cite{bds99,dg01,blp04,g04}, which in turn implies
that $b(p) \neq 1$. On the
other hand, if we suppose that the velocity of the tagged particle obeys a
Markov process and write the corresponding Fokker-Planck equation, in terms of the jump moments, 
$\lim_{\Delta t\to 0}\langle\Delta v\rangle/\Delta t$ and 
$\lim_{\Delta t\to 0}\langle\Delta v^2\rangle/\Delta t$, we obtain
exactly Eq. (\ref{Fokker-Planck}). Here $\langle\dots\rangle$ means average over
different noise (bath) realizations.

\subsection{Coarse grained fields and relevant scales}
We now focus on the study of the hydrodynamic fields of the tagged 
particle with the aid of the Fokker-Planck equation, 
Eq. (\ref{Fokker-Planck}). We define the mean velocity and the temperature of the Brownian particle as
\begin{eqnarray}
\mathbf{u}(t)&=&\int\!\!d\mathbf{r}\!\!\int\!\!d\mathbf{v}\mathbf{v}
F(\mathbf{r},\mathbf{v},t),\\
\frac{d}{2}T(t)&=&\int\!\!d\mathbf{r}\!\!\int\!\!d\mathbf{v}\frac{1}{2}
m(\mathbf{v}-\mathbf{u})^2F(\mathbf{r},\mathbf{v},t).
\end{eqnarray}
Taking moments in the Fokker-Planck equation, we obtain (see Appendix
\ref{appendixB}) 
\begin{eqnarray}\label{ec_u}
\frac{\partial\mathbf{u}(t)}{\partial t}&=&-\gamma_e(t)a(p)\mathbf{u}(t), \\
\label{ec_T}
\frac{\partial T(t)}{\partial t}&=&-2\gamma_e(t)a(p)[T(t)-b(p)T_g(t)].
\end{eqnarray}
As the function $\gamma_e$ is a known functional of the gas density $n_g$ and
temperature $T_g$, equations (\ref{ec_u}) and (\ref{ec_T}) can be
integrated, which yields
\begin{equation}
\mathbf{u}(t)=\mathbf{u}(0)\left[1+\nu_g(0)p(\zeta_n+\zeta_T/2)t\right]
^{-\frac{a(p)\zeta_T}{(2\zeta_n+\zeta_T)\epsilon}},
\end{equation}
and 
\begin{eqnarray}\label{sol_T}
T(t)&=&\frac{b(p)T_g(0)}{1-\epsilon}
\left[1+\nu_g(0)p(\zeta_n+\zeta_T/2)t\right]
^{-\frac{2\zeta_T}{2\zeta_n+\zeta_T}}\nonumber\\
&+&\left[T(0)-\frac{b(p)T_g(0)}{1-\epsilon}\right]
\left[1+\nu_g(0)p(\zeta_n+\zeta_T/2)t\right]
^{-\frac{2\zeta_T}{\epsilon(2\zeta_n+\zeta_T)}}.
\end{eqnarray}
In the above equation, we have introduced the dimensionless coefficient $\epsilon$
\begin{equation}\label{epsilon}
\epsilon=\frac{p\zeta_T\nu_g(t)}{2a(p)\gamma_e(t)}
=\frac{\sqrt{2}d\zeta_T}{2(d+2)a(p)}
\left(\frac{\sigma_g}{\sigma_0}\right)^{d-1}\frac{p}{\Delta},
\end{equation}
that is not necessarily a small quantity.

As can be seen in Eq. (\ref{sol_T}), the behavior of the temperature depends
strongly on the value of $\epsilon$. If $\epsilon<1$ the first term of
equation (\ref{sol_T}) dominates in the long time limit and the 
temperature of the
tagged particle asymptotically decays with the same power as the
temperature of the gas (see equation (\ref{TH_sol})). As a consequence of 
(\ref{sol_T}) we have
\be\label{T_infty}
\lim_{t\to\infty}\frac{T(t)}{T_g(t)}=\frac{b(p)}{1-\epsilon}, 
\qquad \epsilon<1.
\ee
On the other hand, if 
$\epsilon>1$ the second term of equation (\ref{sol_T}) dominates in the 
long time limit and the temperature decays slower than the gas
temperature. One can understand this behavior as follows; the parameter
$\epsilon$ is essentially the quotient between the cooling rate of the gas and
the relaxation rate of the tagged particle's temperature. If the former is
smaller than the latter, the tagged particle's temperature is eventually slaved by
$T_g$ due to the second term of the Eq. (\ref{ec_T}). In the reversed case, the 
tagged particle's temperature evolves independently in the long time limit
with a cooling rate slower than that of the gas. It should be emphasized here that
in the expansion made in the previous section, we implicitly
assumed that $T/T_g$ remains finite, because the coefficients $\mathbf{A}$ and
$N$ were expanded in powers of $\Delta^{1/2}\frac{T}{T_g}$ (see Appendix 
\ref{appendixA}). Hence, the Fokker-Planck equation for $\epsilon>1$ is
restricted to a time window in which $\frac{T}{T_g}$ is small enough. In the
following, we will only consider the case in which the Fokker-Planck equation
is valid for all times, i.e. the double limit  
\begin{equation}\label{limit}
\Delta\to 0, \qquad p\to 0, \qquad \epsilon<1, \qquad
\epsilon\propto\left(\frac{\sigma_g}{\sigma_0}\right)^{d-1}\frac{p}{\Delta},
\end{equation}
where the requirement of small $p$ stems from $\Delta \ll 1$ and $\epsilon$
bounded from above.
Hence, in order to be consistent with this limit, we can substitute in the 
Fokker-Planck equation, Eq. (\ref{Fokker-Planck}), the values
of the coefficients, $a(p)$ and $b(p)$ by their elastic limits 
\begin{equation}
\lim_{p\to 0}a(p)=1, \qquad \lim_{p\to 0}b(p)=1, 
\end{equation}
so that
\begin{equation}\label{Fokker-Planck2}
\left[\frac{\partial}{\partial t}+\mathbf{v}\cdot\nabla\right]
F(\mathbf{r},\mathbf{v},t)=\gamma_e(t)
\frac{\partial}{\partial\mathbf{v}}\cdot
\left[\mathbf{v}+\frac{T_g(t)}{m}
\frac{\partial}{\partial\mathbf{v}}\right]F(\mathbf{r},\mathbf{v},t).
\end{equation}
This equation is formally identical to the one obtained for an elastic gas
\cite{resibois}, except for the fact that the density and temperature of the
gas depend on time. This is a consequence of the elastic limit to which we are
restricted. In general, the coefficients $a(p)$ and $b(p)$, Eqs. (\ref{ec.a})
and (\ref{ec.b}), differ from unity due to the non Maxwellian character of the
distribution function of the bath, and Einstein relation is violated.

In order to analyze this equation, it is convenient to introduce the
dimensionless time scale, $t^*$, proportional to the number of collisions of
the tagged particle
\begin{equation}\label{t^*}
t^*=(1-\epsilon_0)\int_0^tdt^\prime\gamma_e(t^\prime),
\end{equation}
where $\epsilon_0$ is defined by substituting $\zeta_T(p)/a(p)$ in 
(\ref{epsilon}) by its $p\to 0$ limit:
\be
\epsilon_0 =\frac{\sqrt{2}}{16}\left(\frac{\sigma_g}{\sigma_0}\right)^{d-1}\frac{p}{\Delta}.
\ee
The dimensionless time scale $t^*$ is related to the real time $t$ by
\be
t^*=\frac{\gamma_e 2(1-\epsilon_0)}{\nu_g}\frac{1}{p (2\zeta_n + \zeta_T)}\log[1+\nu_g(0)p(\zeta_n+\zeta_T/2)t].
\ee
In this time scale, the
evolution of the mean velocity and temperature of the tagged particle are
particularly simple
\be\label{ec:u_est}
\mathbf{u}(t^*)=\mathbf{u}(0)e^{-\frac{t^*}{1-\epsilon_0}},
\ee
and 
\be\label{ec:T_est}
\frac{T(t^*)}{T_g(t^*)}
=\frac{T(0)}{T_g(0)}e^{-2t^*}+\frac{1}{1-\epsilon_0}(1-e^{-2t^*}). 
\ee
Such predictions will be compared against numerical simulations in section 
\ref{simulations}.
Let us also introduce the scaled distribution
\begin{equation}
F(\mathbf{r},\mathbf{v},t)=\frac{1}{v_\epsilon^d(t)}
F^*(\mathbf{r},\mathbf{v}^*,t^*), \qquad 
\mathbf{v}^*=\frac{\mathbf{v}}{v_\epsilon(t)},
\end{equation}
where
\begin{equation}\label{v_epsilon}
v_\epsilon(t)=\left(\frac{1}{1-\epsilon_0}\right)^{1/2}
\left(\frac{2T_g(t)}{m}\right)^{1/2}.
\end{equation}
The function $v_\epsilon(t)$ is introduced because with these definitions we
have 
\begin{equation}
v_\epsilon(t)\to\left[\frac{2T(t)}{m}\right]^{1/2}, 
\end{equation}
in the long time limit. 
In these variables the Fokker-Planck equation (\ref{Fokker-Planck2}) reduces to
\begin{equation}\label{Fokker-Planck3}
\left(\frac{\partial}{\partial t^*}+\ell_0(t^*)\mathbf{v}^*\cdot\nabla\right)
F^*(\mathbf{r},\mathbf{v}^*,t^*)=\Lambda_{FP}(v^*)
F^*(\mathbf{r},\mathbf{v}^*,t^*),
\end{equation}
where we have introduced the standard homogeneous Fokker-Planck operator
\be
\Lambda_{FP}(v^*)=\frac{\partial}{\partial\mathbf{v}^*}\cdot
\left(\mathbf{v}^*+\frac{1}{2}\frac{\partial}{\partial\mathbf{v}^*}\right),
\ee
and the function proportional to the mean free path 
\begin{equation}
\ell_0(t^*)=\frac{v_\epsilon(t)}{(1-\epsilon_0)\gamma_e(t)}=
\frac{d\Gamma(d/2)}{(1-\epsilon_0)^{3/2}4\pi^{\frac{d-1}{2}}}\Delta^{-1/2}
\left(\frac{\sigma_g}{\sigma_0}\right)^{d-1}[n_g(t^*)\sigma_g^{d-1}]^{-1}.
\end{equation}
Taking into account the definition of $\epsilon$, Eq. (\ref{epsilon}), we can
write explicitly  $\ell_0$ as a function of the $t^*$ variable as
\be\label{ell_0}
\ell_0(t^*)=\ell_0(0)e^{\epsilon^*t^*}, \qquad 
\epsilon^*=\frac{2\zeta_n\epsilon_0}{\zeta_T(1-\epsilon_0)}.
\ee

To sum up, we have obtained the evolution equation for the distribution
function of a tagged particle in a bath of particles which annihilate, in the
limit where the tagged particle is much heavier than the particles of the
bath. There are some points in common with the elastic case, but also some
important differences. The homogeneous operator, whose spectral properties
are well-known \cite{resibois,mclennan}, is exactly the same, but the flux
term is weighted by a function depending on time and that diverges in the long
time limit. This will have important consequences in the study of diffusion as
we will see in the following section. Moreover, the equation is not valid for
all values of the probability $p$ of annihilation in the bath and 
masses of the tagged particle but, as already mentioned, is limited to the
double limit of Eq. (\ref{limit}), in which $\epsilon_0<1$.
\end{section}

\begin{section}{Long-time limit solution of the Fokker-Planck equation}

In this section we investigate the long time behavior of the solution of the
Fokker-Planck equation, Eq. (\ref{Fokker-Planck3}), starting with an arbitrary 
initial condition. The 
objective is to study if the tagged particle reaches some scaling state
in the long time limit and also to analyze how the particle diffuses.

\subsection{Evolution towards a scaling form}
 As the Fokker-Planck equation is linear, it is convenient to work in the Fourier
space. The Fourier component of the tagged particle distribution function is
defined as
\be
F_\mathbf{k}(\mathbf{v}^*,t^*)=\int\!\!d\mathbf{r}
e^{-i\mathbf{k}\cdot\mathbf{r}}F^*(\mathbf{r},\mathbf{v}^*,t^*),
\ee
so that Eq. (\ref{Fokker-Planck3}) yields
$F_\mathbf{k}(\mathbf{v}^*,t^*)$
\begin{equation}\label{F-P-F}
\frac{\partial}{\partial t^*}F_\mathbf{k}(\mathbf{v}^*,t^*)
=[\Lambda_{FP}(v^*)-i\ell_0(t^*)\mathbf{k}\cdot\mathbf{v}^*]
F_\mathbf{k}(\mathbf{v}^*,t^*).
\end{equation}
The spectrum of the operator 
$\Lambda_{FP}(v^*)-i\ell_0(t^*)\mathbf{k}\cdot\mathbf{v}^*$ is  known
\cite{resibois,mclennan}. The eigenvalues are 
\be\label{eigenvalues}
\lambda_\mathbf{n}(\mathbf{k},t)=-\frac{1}{2}[k\ell_0(t)]^2-\sum_{j=1}^dn_j, 
\ee
where we have introduced the vector label $\mathbf{n}=(n_1,\dots,n_d)$, with
possible coordinate values $n_i=0,1,2,\dots\infty$. Hence, for any initial condition,
all the $\mathbf{k}$-Fourier components decay and only the $\mathbf{k}=\mathbf{0}$ 
remains. Moreover, as the eigenfunction associated to the vanishing eigenvalue is the Maxwellian distribution \cite{resibois,mclennan}
\begin{equation}
\chi_M(\mathbf{v}^*)=\frac{1}{\pi^{d/2}}e^{-v^{*2}},
\end{equation}
we obtain
\begin{equation}
F(\mathbf{r},\mathbf{v},t)\to \frac{1}{v_\epsilon^d(t)}\chi_M{(v^*)}, 
\end{equation}
in the long time limit. 

As a consequence, the tagged particle distribution function approaches a
scaling form similar to (\ref{scalingHDS}) 
for the gas, but with a different temperature (see
(\ref{T_infty})). In this regime the cooling
rates of the bath and of the tagged particle are the same and the temperatures
are proportional. The situation is similar to that of an elastic particle
in a bath of inelastic grains \cite{bds99,brgd99}. Nevertheless, there is
an important difference: in the inelastic case it has been proved that there
exists an exact mapping with an elastic system. On the other hand, in our
problem such a mapping fails due to the flux term which
explicitly depends on time. 

\subsection{Characteristics of diffusive motion}

Our objective is to study the evolution
equation for the density of tagged particles, 
$n(\mathbf{r},t)=\int d\mathbf{v}F(\mathbf{r},\mathbf{v},t)$ in a
``macroscopic'' scale, i.e. in a long time and length scale compared to the 
microscopic ones. The
microscopic time scale is defined by the slowest kinetic modes of 
$\Lambda_{FP}$, i.e. the modes with a single non vanishing component,
labeled by $n_i=\delta_{ij}$ for a given 
value of $j$ in $[1,d]$. The microscopic
length scale is defined by the mean free path of the tagged particle which is
proportional to $\ell_0(t^*)$. The starting point will be the Fokker-Planck
equation for $F_\mathbf{k}$, Eq. (\ref{F-P-F}). As the generator of the 
dynamics, the operator 
$\Lambda_{FP}(v^*)-i\ell_0(t^*)\mathbf{k}\cdot\mathbf{v}^*$, does not commute
with its time derivative, it is not possible to write the general solution of
equation (\ref{F-P-F}) in terms of the initial condition in a simple
way. Nevertheless, it is shown in Appendix \ref{appendixC} that in the
hydrodynamic limit, i.e. $k\ell_0(t^*) \ll 1$ for all the time evolution and 
$t^* \gg 1$, a closed equation for the
Fourier component of the density, 
$n_\mathbf{k}=\int d\mathbf{v}^*F_\mathbf{k}(\mathbf{v}^*,t^*)$, is obtained
\be\label{ec:n_k1}
\frac{\partial n_\mathbf{k}(t^*)}{\partial t^*}=-D_0[k\ell_0(t^*)]^2
n_\mathbf{k}(t^*), 
\ee
where the diffusion coefficient is
\be
D_0=\frac{1}{2(1+\epsilon^*)}.
\ee
This asymptotic behavior can be evaluated, taking advantage of the
scale separation (i.e. the mode with $\mathbf{n}=\mathbf{0}$ is 
isolated from the other modes). From equation (\ref{ec:n_k1}) we can derive the
evolution equation for the density
\be\label{ec:nr}
\frac{\partial n(\mathbf{r},t^*)}{\partial t^*}=D_0\ell_0^2(t^*)
\nabla^2 n(\mathbf{r},t^*).
\ee
This is the equation we were looking for and it is only valid in the
``macroscopic'' time and length scale. If we transform this equation to real
time with the aid of formula (\ref{t^*}) we obtain
\be
\frac{\partial n(\mathbf{r},t)}{\partial t}=D(t)
\nabla^2 n(\mathbf{r},t), \qquad D(t)=
\frac{\zeta_T}{(1-\epsilon_0)[\zeta_T+(2\zeta_n-\zeta_T)\epsilon_0]}D_e(t),
\ee
where $D_e(t)=2v_g^2(t)/\gamma_e(t)$ is the same as the diffusion coefficient
for elastic collisions except that it appears here 
as a function of the time-dependent gas
temperature and density. As can be seen, for our system the diffusion
coefficient $D(t)$ is far from being a trivial generalization of the elastic
diffusion coefficient. 

Now let us focus on the predictions of our diffusion equation. To this end,
we introduce the mean square 
displacement
\begin{equation}
\langle r^2(t^*)\rangle=\int\! d\mathbf{r}\,r^2 n(\mathbf{r},t^*).
\end{equation}
If we consider an infinite system, we obtain from equation (\ref{ec:nr}) 
\begin{equation}\label{ec:rcm}
  \frac{\partial}{\partial t^*}\langle r^2(t^*)\rangle=2dD_0 \ell_0(t^*)^2,
\end{equation}
that will only be valid in the long time limit. It is straightforward to integrate
equation (\ref{ec:rcm}), taking into account the explicit formula for 
$\ell_0(t^*)$, Eq. (\ref{ell_0}). This gives
\be\label{diffusion1}
\langle  r^2(t^*)\rangle=dD_0\ell_0^2(0)
\frac{e^{2\varepsilon^*t^*}-1}{\varepsilon^*}, 
\ee
or in real time
\be
\langle  r^2(t)\rangle=\frac{dD_0\ell_0^2(0)}{\epsilon^*}
\left\{\left[1+\nu_g(0)p(\zeta_n+\zeta_T/2)t\right]
^{\frac{4\zeta_n}{2\zeta_n+\zeta_T}}-1\right\}.
\ee
As can be seen in Eq. (\ref{diffusion1}), the diffusive behavior is completely
different from its elastic  or even  inelastic counterparts,
where it was found that the mean square displacement is proportional
to the number of collision per particle \cite{bds99,brgd99}. The fact that the 
bath is loosing particles significantly affects this dynamics, and the mean square
displacement increases exponentially in the collision per particle scale. As we
will see in the following section, simulation results agree well with
our theoretical prediction. Roughly speaking, the exponent 
$4\zeta_n/(2\zeta_n+\zeta_T)$ is close to $8d/(4d+1)$
(approximately 1.77 in two dimensions, and 1.84 in three dimensions).

\subsection{Diffusive behavior : an alternative derivation}
In the remainder of this section, we show that, under plausible
hypothesis, it is possible to re-derive ``\`a la Einstein'' the formula for the
mean square displacement, Eq. (\ref{diffusion1}). This derivation has the
merit of leading to a Green-Kubo like expression for the 
diffusion coefficient. We start by writing the mean-squared displacement as
\be
\langle  r^2(t)\rangle=\int_0^tdt'\int_0^tdt''\langle\mathbf{V}(t')\cdot
\mathbf{V}(t'')\rangle. 
\ee
Here, the position and the velocity of the tagged particle, $\mathbf{r}(t)$
and $\mathbf{V}(t)$, are considered as a stochastic process and
$\langle\dots\rangle$ denotes an ensemble average over different
trajectories. 
Let us change the variables from
$t\to t^*$ and let us also introduce the scaled velocity
\be
\mathbf{w}(t^*)\equiv\frac{\mathbf{V}}{v_\epsilon(t)}, 
\ee
where $v_\epsilon(t)$ is defined in (\ref{v_epsilon}). With these definitions
we have
\begin{eqnarray}
\langle r^2(t^*)\rangle&=&\frac{1}{(1-\epsilon_0)^2}
\int_0^{t^*}ds_1\gamma^{-1}_e(s_1)\int_0^{t^*}ds_2\gamma^{-1}_e(s_2)
\langle\mathbf{V}(s_1)\cdot\mathbf{V}(s_2)\rangle\nonumber\\
&=&\ell_0^2(0)\int_0^{t^*}ds_1\int_0^{t^*}ds_2e^{\epsilon^*(s_1+s_2)}
\langle\mathbf{w}(s_1)\cdot\mathbf{w}(s_2)\rangle, 
\end{eqnarray}
where we have used the definitions of $t^*$ and $\ell_0(0)$, Eq. (\ref{t^*})
and Eq. (\ref{ell_0}). Now, if we assume that the tagged particle is in the
scaled regime, i.e. the temperature is $T_g(t^*)/(1-\epsilon_0)$ and that the
correlation function $\langle\mathbf{w}(s_1)\cdot\mathbf{w}(s_2)\rangle$ is a
function of $s_1-s_2$, by integrating in the new variables
$S=(s_1+s_2)/2$ and $s=s_1-s_2$, the following relation is obtained
\be\label{r2}
\frac{\langle r^2(t^*)\rangle}{e^{2\epsilon^*t^*}}
\frac{\epsilon^*}{d\ell_0^2(0)}\to\frac{1}{d}\int_0^{t^*} ds
\langle\mathbf{w}(s)\cdot\mathbf{w}(0)\rangle e^{-\epsilon^*s}, 
\ee
in the long time limit. 
This formula is the generalization of the Einstein formula for the diffusion
coefficient of a heavy particle in a fluid in the homogeneous decay state. It relates the asymptotic
behavior of the mean-square displacement with the time integral of the
autocorrelation function of the velocity weighted by the exponential
$e^{-\epsilon^*s}$. So far we have considered the Fokker-Planck equation as
the equation for the one-time probability distribution function. If we assume 
that the velocity of the tagged particle is a Markov process, then the
Fokker-Planck equation is also the equation for the conditional probability
and we can evaluate easily the correlation function 
$\langle\mathbf{w}(s_1)\cdot\mathbf{w}(s_2)\rangle$. Taking into account Eqs. 
(\ref{ec:u_est}) and (\ref{v_epsilon}), we have
\be\label{cor_ww}
\langle\mathbf{w}(s_1)\cdot\mathbf{w}(s_2)\rangle=\frac{d}{2}e^{-|s_1-s_2|}.
\ee
By substituting this formula into Eq. (\ref{r2}) we re-derive
Eq. (\ref{diffusion1}) with the same diffusion coefficient $D_0$, that can be
written in the Green-Kubo form
\be
D_0=\frac{1}{d}\int_0^\infty ds
\langle\mathbf{w}(s)\cdot\mathbf{w}(0)\rangle e^{-\epsilon^*s}.
\ee
\end{section}

\begin{figure}
\begin{minipage}[c]{1.0\textwidth}
\begin{center}
  \includegraphics[angle=0,width=0.4\textwidth]
{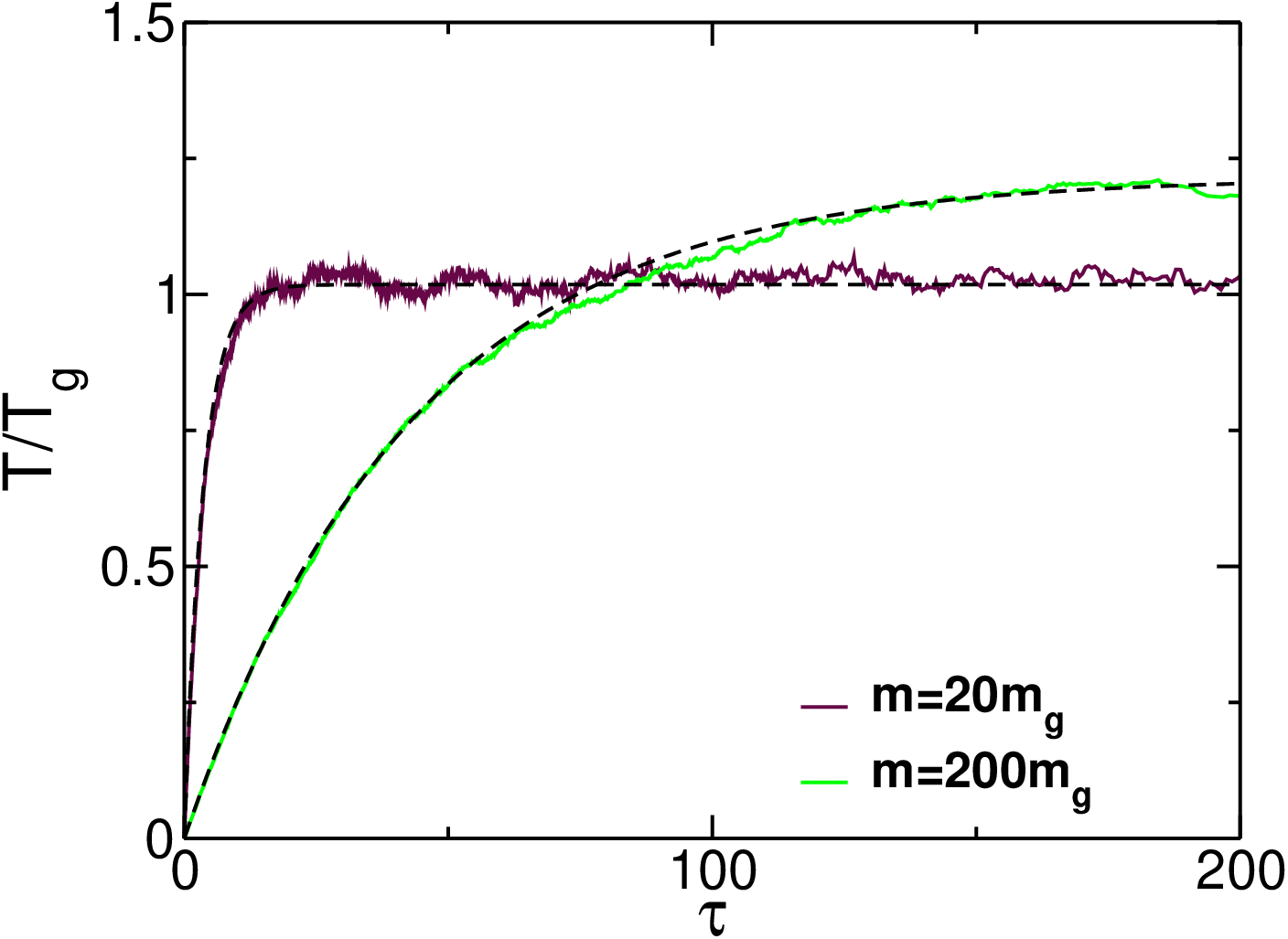}
\end{center}
\end{minipage}
\caption{ (Color online) Evolution of the temperature ratio  as a function of the 
number of collisions per particle $\tau$ for a system with $p=0.01$ and two 
values of the tagged particle's mass, $m=20 m_g$ ($\Delta =1/20$) and $m=200 m_g$ ($\Delta = 1/200$). 
The dashed line is
the theoretical prediction given by Eq. (\ref{ec:T_est}).}\label{figura0}
\end{figure}

\begin{section}{Direct Simulation Monte Carlo results}\label{simulations}

The objective of this section is to put to the test the main results of the previous sections by means of the direct simulation Monte Carlo method (DSMC).
More precisely, we will analyze the temperature and mean velocity evolution, together with  
the tagged particle diffusion.  
We have performed DSMC simulations of a system of $N_g$ hard disks of mass $m_g$ 
and diameter $\sigma_g$ which annihilate with probability $p$ or collide 
elastically with probability $1-p$ everytime two particles meet. 
Bird's algorithm \cite{bird} has been used. The parameters in all the simulations 
were $m_g=1$, $\sigma_g=1$, $N_g(0)=10^5$ and 
$T_g(0)=1$. We have considered only one tagged particle in each simulation,
that collides elastically with the surrounding bath. The diameter of this
particle has been set to unity ($\sigma=1$) and we have varied the value of its mass $m$. The 
values of the probability of annihilation $p$ of the particles in the bath have
been 
$p=0.1$ and $p=0.01$, and the results have been averaged over $2\cdot 10^4$ 
and $4\cdot 10^3$ trajectories respectively.
For a given value of $p$, we have performed a series of simulations for
different values of the mass of the tagged particle. Taking due account of
the constraint $\epsilon_0<1$, the value of the tagged 
particle's mass must be smaller than $10^2$ for $p=0.1$ and $10^3$ for 
$p=0.01$.   
\begin{figure}
\begin{center}
\includegraphics[angle=0,width=0.4\textwidth]{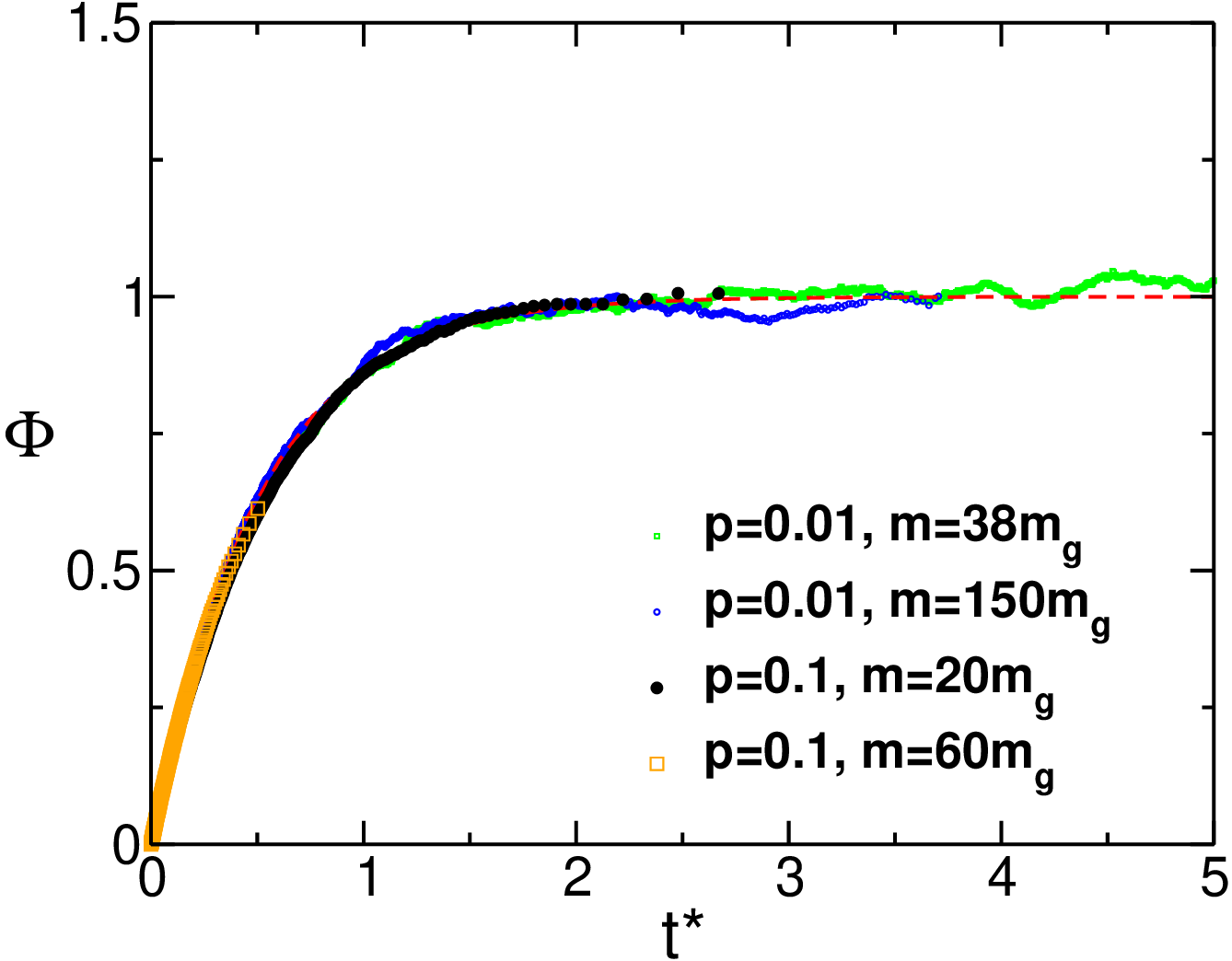}
\end{center}
\caption{ (Color online) Scaling function $\Phi$ defined in the main text as a function of reduced time 
$t^*$ for different
systems with $p=0.1$ and $p=0.01$ and different values of the tagged
particle's mass. The dashed line is the theoretical prediction of Eq. 
(\ref{eq:100}). }
\label{figura2}
\end{figure}
\begin{figure}
\begin{center}
\includegraphics[angle=0,width=0.4\textwidth]{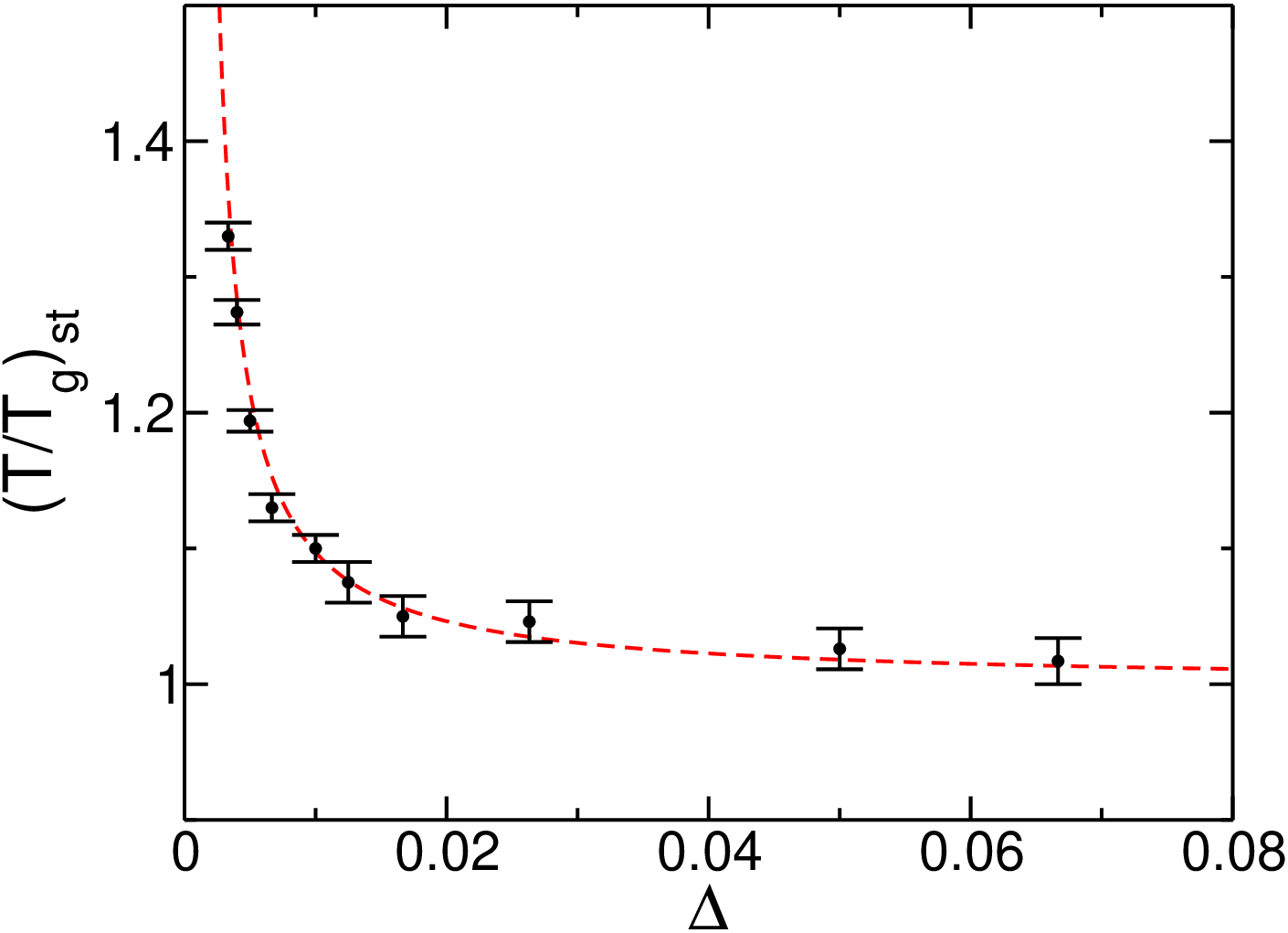}
\end{center}
\caption{ (Color online) Stationary values of the temperatures ratio for a system with 
$p=0.01$ and different values of the tagged particle's mass $\Delta=m_g/m$. 
Symbols are for the Monte Carlo data and the dashed line shows
the long time limit of Eq. 
(\ref{ec:T_est}).}
\label{figura2bis}
\end{figure}

\begin{figure}
\begin{center}
\includegraphics[angle=0,width=0.4\textwidth]{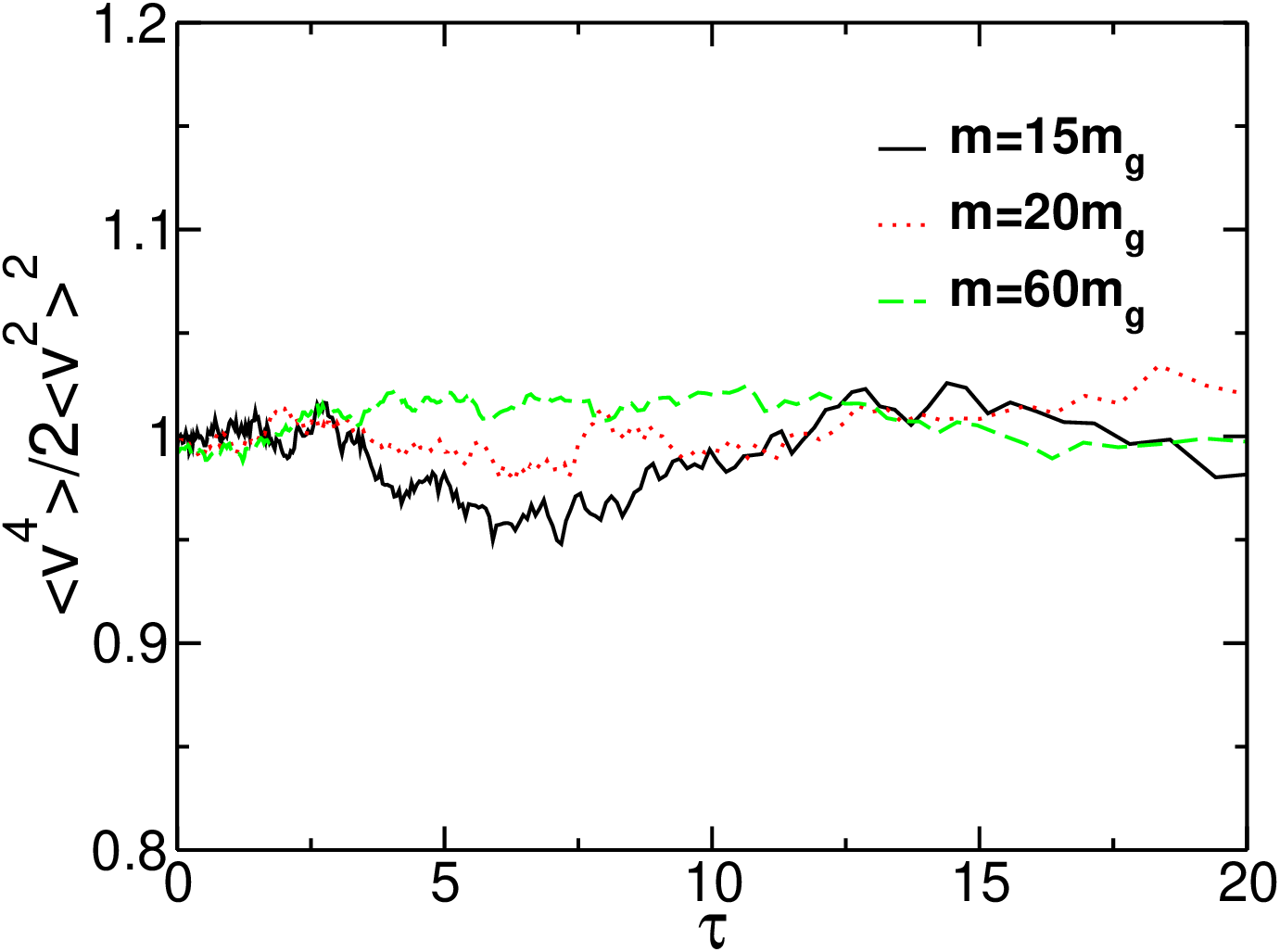}
\end{center}
\caption{ (Color online) Reduced fourth moment of the tagged particle velocity as a function
  of the dimensionless time $\tau$ for $p=0.1$ and different values of the
  tagged particle's mass.}
\label{fig2con5}
\end{figure}

Figure \ref{figura0} shows the 
evolution of the ratio 
$T/T_g$ for a system  with $p=0.01$ and for 
two values of the tagged particle's mass, that is, $m=20m_g$ and $m=200m_g$, as a function of the number of collisions per particle, $\tau$, 
defined as
\be\label{escalatau}
\tau=\frac{1}{2}\int_0^tdt^\prime \nu_g(t^\prime)=\frac{1}{2(1-\epsilon_0)}\frac{\nu_g}{\gamma_e} t^*. 
\ee
The initial value of temperature of the tagged particle is $T(0)=0$ for all the 
trajectories, since at $t=0$, the intruder has a prescribed velocity. 
As we can see, in this scale, the evolution to the stationary
value of the ratio of the temperatures is slower as we increase the mass of
the tagged particle. This implies that there are values of the tagged
particle's mass for which the ratio of the temperatures will not reach its 
stationary value in the time of the simulation (for instance, 
the number of particles in the 
bath for $\tau=200$ is $N_g\simeq 1800$, which hinders correct statistical sampling). 
The theoretical prediction in this time 
scale (dashed line) is obtained directly from Eq. (\ref{ec:T_est}), taking 
into account Eq. (\ref{escalatau}). The agreement between theory and 
simulations is good. 

Similar simulations were performed with different values of the tagged particle mass
($m$ ranging from 15 to 100 for $p=0.1$ and from 15 to 900 for $p=10^{-2}$). Since 
$T_g(0)=0$, Eq. (\ref{ec:T_est}) predicts 
\be\label{eq:100}
\Phi \, \equiv \, (1-\epsilon_0) \frac{T}{T_g} \, = \, 1-e^{-2 t^*}.
\ee
As can be observed in Fig. \ref{figura2}, all simulation data for $\Phi$ collapse 
onto a single curve.
The time scale
$t^*$ can be calculated from the scale $\tau$ defined in (\ref{escalatau}).
In the same vein, we can obtain the stationary value of the temperature ratio, which is
plotted as a function of $\Delta = m_g/m$ in Fig. \ref{figura2bis}~; this validates
our theoretical analysis.

\begin{figure}
\begin{minipage}[c]{1.0\textwidth}
\begin{center}
\includegraphics[angle=0,width=0.4\textwidth]{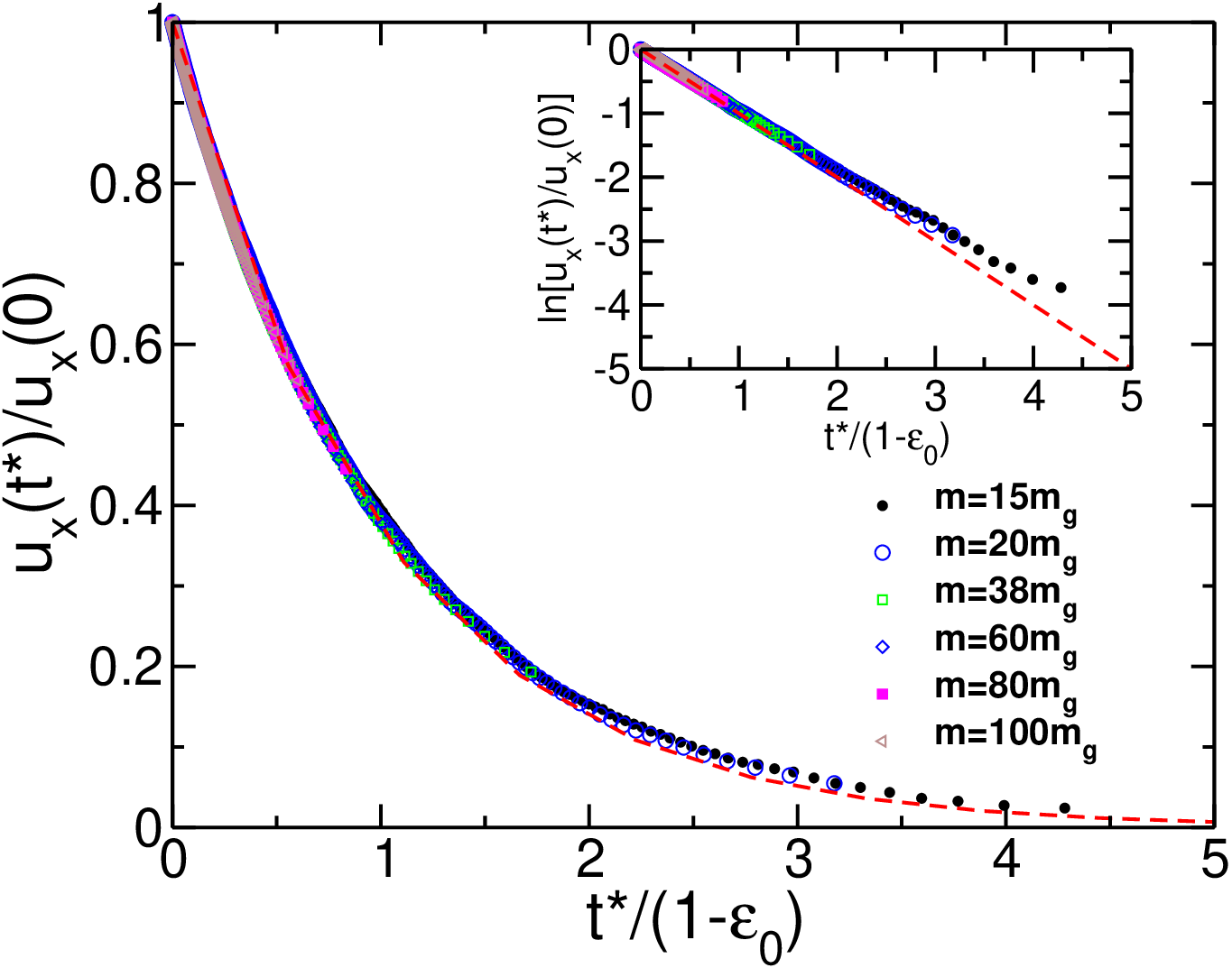}
\end{center}
\end{minipage}
\caption{ (Color online) Mean velocity of the tagged particle as a function of 
$t^*/(1-\epsilon_0)$ for a system with $p=0.1$ and several values of the 
tagged particle's mass. The dashed line is the 
theoretical prediction given by Eq. (\ref{ec:u_est}). Inset: Mean velocity of 
the tagged particle as a function of 
$t^*/(1-\epsilon_0)$ for a system with $p=0.1$ on a logarithmic scale. }
\label{figura3}
\end{figure}

In order to probe --at least partially-- the Gaussian nature of the time dependent tagged particle
velocity statistics, we have measured the reduced 
fourth moment $4\langle v^4\rangle/(d(d+2)\langle v^2\rangle^2)$. 
As can be seen in Fig. \ref{fig2con5}, 
where we have plotted the results for $p=0.1$ and several values of $\Delta$ 
as a function of the number of collisions per particle, $\tau$, the value of 
this quantity is in agreement with the Gaussian prediction (that is unity)
within the statistical uncertainties. 

Consider next the mean velocity of the tagged particle, $\mathbf{u}(t^*)$.
In order to study the decay of this quantity, we have 
performed a set of simulations starting with a component of the velocity field
in the $x$ direction when it is
immersed in a bath in the homogeneous decay state, $u_x(0)$. 
In Fig. \ref{figura3}, we plot $u_x(t^*)/u_x(0)$ as a function of 
$t^*/(1-\epsilon_0)$ for $p=0.1$ and several values of $m$. In this time scale,
the data for all values of $m$ collapse due to Eq. (\ref{ec:u_est}). In 
the inset, the same quantity is plotted on a logarithmic 
scale. If the theoretical prediction in Eq. (\ref{ec:u_est}) is verified, the 
above plot must lead to a straight line with slope $\zeta_\mathbf{u}=1$ 
(dashed line), where $\zeta_\mathbf{u}$ is the decay rate of the mean 
velocity. On the other hand, $\zeta_\mathbf{u}$ can be fitted on the logarithmic plot
of the inset. Reporting the corresponding measures in Fig. 
\ref{figura4} against the mass ratio, it appears that the theoretical prediction
$\zeta_\mathbf{u}=1$ is approached as $\Delta\to 0$, as expected.

\begin{figure}
\begin{minipage}[c]{1.0\textwidth}
\begin{center}
\includegraphics[angle=0,width=0.4\textwidth]
{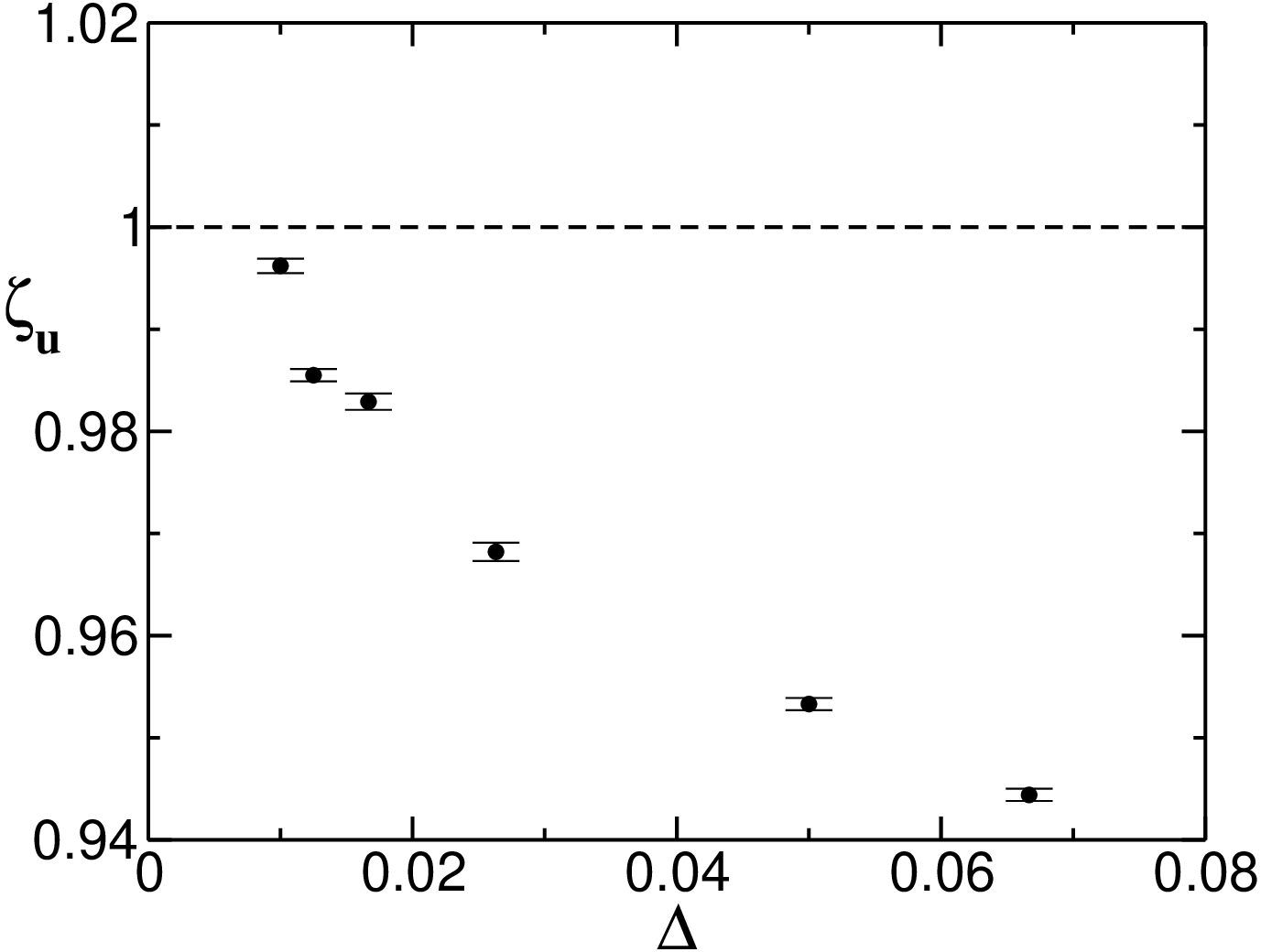}
\end{center}
\end{minipage}
\caption{Decay rate of the mean velocity as a function of $\Delta$ for a 
system with $p=0.1$. The symbols are from DSMC simulations an the dashed line 
is the theoretical prediction. }\label{figura4}
\end{figure}

\begin{figure}
\begin{minipage}[c]{1.0\textwidth}
\begin{center}
  \includegraphics[angle=0,width=0.4\textwidth]
{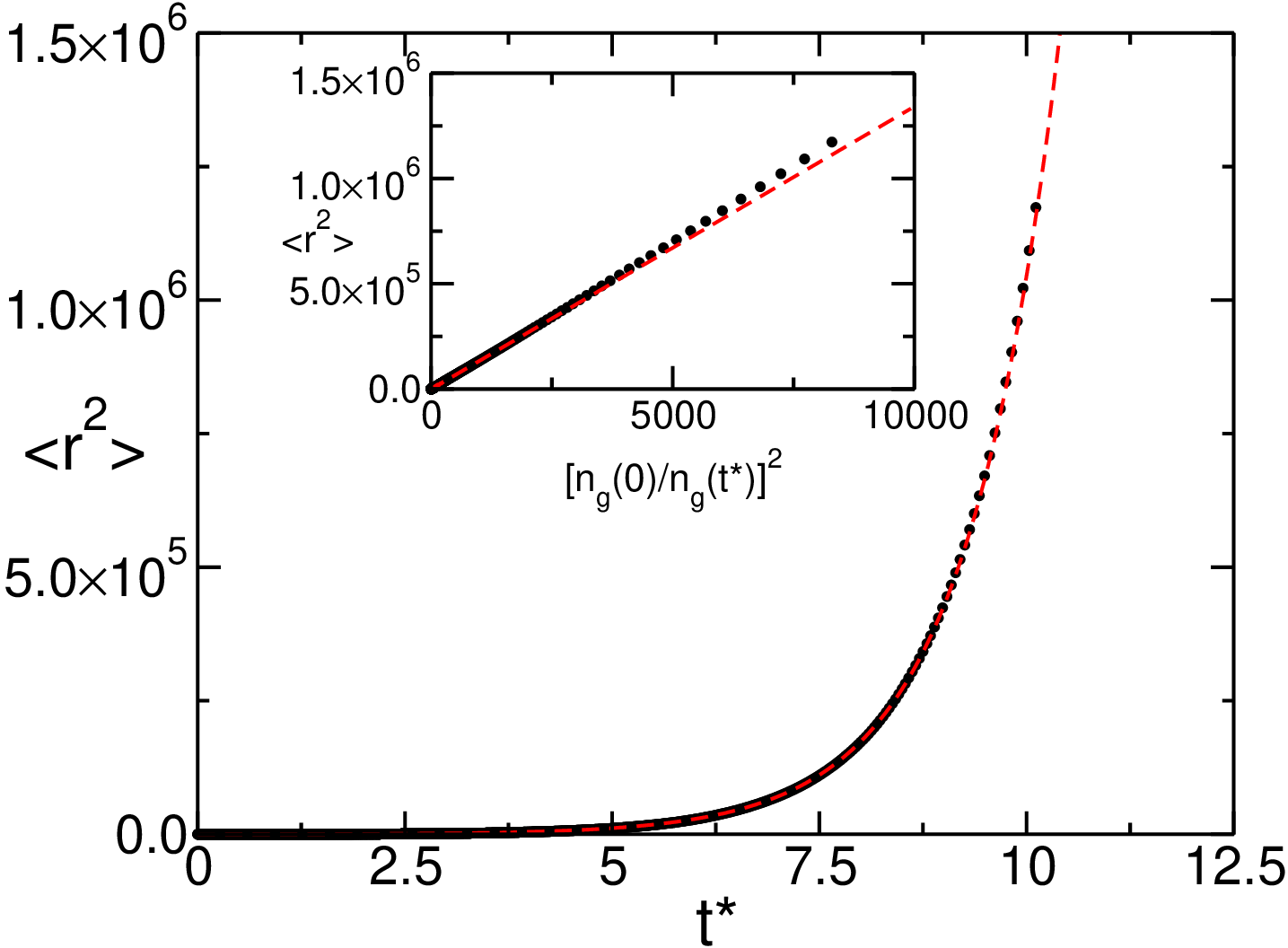}
\end{center}
\end{minipage}
\caption{ (Color online) Mean squared displacement for a system with $p=0.01$ and $m=60m_g$ as a function of $t^*$. The dashed line is the theoretical prediction, Eq. (\ref{diffusion1}). Inset : same quantity as a function of 
$n_g(0)^2/n_g(t^*)^2$. The dashed line is the theoretical prediction of  Eq. (\ref{lenfuncN})
where $B$ follows from (\ref{eq:200}).}
\label{figura6}
\end{figure}

\begin{figure}
\begin{minipage}[c]{1.0\textwidth}
\begin{center}
  \includegraphics[angle=0,width=0.4\textwidth]
{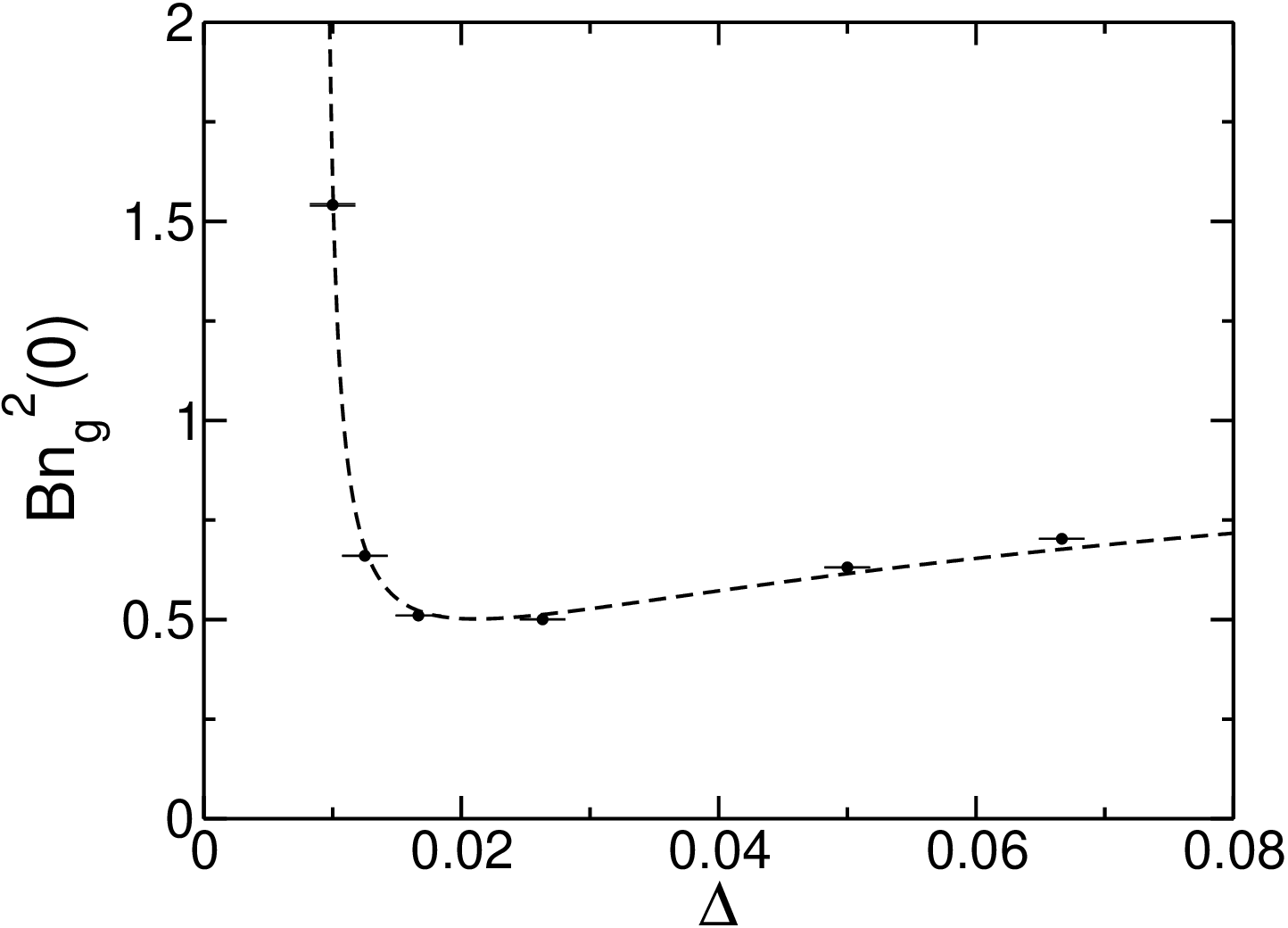}
  \includegraphics[angle=0,width=0.4\textwidth]
{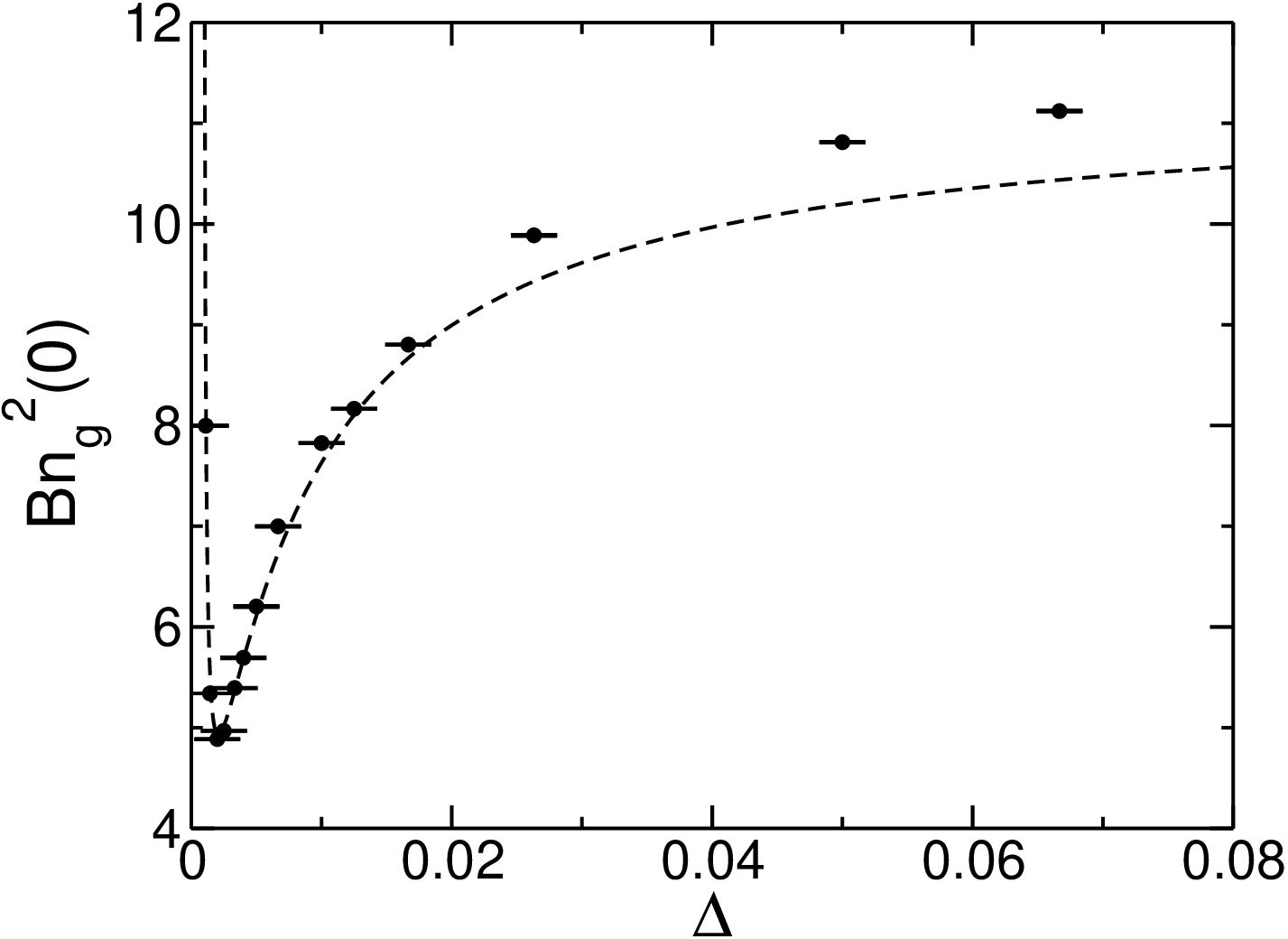}
\end{center}
\end{minipage}
\caption{Values of $Bn_g^2(0)$ as a function of $\Delta$ for systems with 
$p=0.1$ (left) and $p=0.01$ (right). The dashed line is for the prediction of
Eq. (\ref{eq:200}).}
\label{figura7}
\end{figure}

Finally, the accuracy of the prediction for the diffusion equation has also 
been tested by measuring in DSMC simulations the mean square displacement of the
tagged particle. In contrast to the granular case phenomenology and as a 
consequence of the continuous decay of particles in the bath, the mean 
squared displacement increases exponentially in the $t^*$ scale, see Eq. 
(\ref{diffusion1}). In Fig. \ref{figura6}  we have plotted the time 
evolution of $\langle r^2\rangle$ in the scale $t^*$ for a system with 
$p=0.01$ and $m=60 m_g$. The dashed line is the theoretical prediction given by 
Eq. (\ref{diffusion1}), and shows good agreement with numerical data. The same quantity, 
written in terms of the bath density, reads
\begin{eqnarray}\label{lenfuncN}
\langle  r^2(t^*)\rangle&=&dD_0\ell_0^2(0)
\frac{e^{2\epsilon^*t^*}-1}{\epsilon^*}\nonumber\\
&=&B\left[\frac{n_g(0)^2}{n_g(t^*)^2}-1\right],
\end{eqnarray}
where we have defined 
\begin{equation}
B=\frac{dD_0\ell_0^2(0)}{\varepsilon^*}, 
\label{eq:BB}
\end{equation}
and we have taken into account that $n_g(t^*)=n_g(0)e^{-\epsilon^*t^*}$. It then
appears that the mean squared displacement increases linearly with
$n_g(0)^2/n_g(t^*)^2$. This is full agreement with the simulation results,
see the inset of Fig \ref{figura6}. Such a plot allows us 
to extract by linear fitting  the coefficient $B$, which can then
be compared against the prediction of Eq. (\ref{eq:BB}), which explicitly reads
\be\label{eq:200}
B =-\frac{16\sqrt{2}d^2\pi^{1-d}\sigma_g^{2(1-d)}\Gamma\left(\frac{d}{2}\right)^2}{n_g^2(0)p(-16+\sqrt{2}\frac{p}{\Delta})(16+\sqrt{2}(4d-1)\frac{p}{\Delta})},
\ee
where we have taken into account that $\sigma_0=\sigma_g$.
Such a comparison is worked out in Figure \ref{figura7} and fully
corroborates the theoretical analysis, with again an improved agreement
when  $\Delta$ decreases.

It would be also interesting to confirm our theoretical predictions with
Molecular Dynamics simulations. In the low density limit, it is expected to
get similar results. In fact, some simulations were performed finding
qualitatively the same behavior but with much more statistical inaccuracies.

\end{section}

\begin{section}{Conclusions}

In this paper, the diffusive behavior of a tagged intruder immersed in a gas
of particles undergoing ballistic annihilation (i.e. which annihilate with
probability $p$ or  scatter elastically otherwise), has been analyzed. The
collisions between the tagged particle and the surrounding gas are
elastic. Some similarities are found  between our system and the elastic or
inelastic case \cite{resibois,bds99}, but, on the other hand, important
differences arise as a consequence of the continuous decay of particle number
in the system. 

We start from the Boltzmann-Lorentz equation for the distribution function of 
the tagged particle, which is valid, in principle, for arbitrary mass of the
tagged particle. In the limit of a very massive tagged particle, a 
Fokker-Planck equation for the distribution function is derived by means of a
systematic expansion in the mass ratio $\Delta$. Our approach holds 
in the limit $\Delta \ll 1$, but we additionally have the more stringent
condition that the parameter introduced in Eq. 
(\ref{epsilon}), $\epsilon\propto\frac{p}{\Delta}$, must be smaller 
than unity. Analysis of the Fokker-Planck equation leads to predictions for
the temperature ratio, the decay rate of the mean velocity of the tagged
particle and the diffusion coefficient. As in the inelastic case \cite{bds99},
the theory predicts that the ratio between the temperatures of the tagged
particle and the gas is constant in the long time limit as a consequence of
equilibrating cooling rates. When represented in the appropriate time scale,
which is proportional to the number of collisions experienced by the tagged
particle, temperature ratios collapse for all  values of $\Delta$ and $p$
considered. Likewise, the mean intruder velocity (averaged over bath
realizations) decays exponentially. The dynamics of the distribution function
of the tagged particle is governed by a Fokker-Planck operator which spectral
properties are known. More 
specifically, as the eigenvalues of this operator are non-positive, the 
distribution of the tagged particle approaches a Gaussian in the long-time 
limit and admits a scaling form similar to the one for the distribution 
function for the particles in the gas but with a different temperature. 
At variance with the situation of an 
elastic intruder in a bath 
of inelastic particles \cite{sd06}, there is apparently no mapping between our problem and a well chosen 
elastic system. A unique vanishing eigenvalue is responsible for the
slow diffusive behavior of the tagged particle density.
The corresponding diffusion equation has been derived in the hydrodynamic limit, by means of 
a projector decomposition, which yields an explicit expression for the diffusion
coefficient. From a different point of view, the expression for the mean 
squared displacement has also been derived ``\`a la Einstein''. Following this route,
the diffusion coefficient is expressed as a Green-Kubo formula in terms of a weighted time
integral of the tagged particle velocity correlation function. 
This provides a more physical perspective on the results 
derived from the projector method. As already mentioned, the mean squared 
displacement for this system does not increase linearly in the 
collision per particle time scale, as is the case in the elastic and inelastic
cases. This different  behavior is due to the time dependent bath density. 
In the elastic case, both the temperature and the 
density do not depend on time. In an inelastic system \cite{bds99,brgd99}, the time dependence 
goes through the temperature and could be absorbed in the collision per 
particle time scale, which turns out to be impossible in our system
where the mean free path 
$\ell(t^*)$ is an increasing function of time.

Finally, our analytical results have been tested by numerical simulations, and a
very good agreement has been reported for all the range of parameters considered. 
As expected, the agreement is all the better as $\Delta$ is smaller.
In summary, the work reported here provides an example of the accuracy of 
hydrodynamics to describe a system in which there are 
no conserved quantities  in binary encounters (no collisional invariants).

\end{section}

\begin{acknowledgments}
We acknowledge useful discussions with G. Schehr and A. Barrat.
We would like to thank the Agence Nationale de la Recherche
for financial support (grant ANR-05-JCJC-44482). 
M.~I.~G.~S. and P.~M. acknowledge financial support from Becas de
la
Fundaci\'on La Caixa y el Gobierno Franc\'es. M.~I.~G.~S. would like to thank the HPC-EUROPA project (RII3-CT-2003-506079),
with the support of the European Community Research Infrastructure Action, for financial support.

\end{acknowledgments}

\appendix

\begin{section}{Some useful approximations}
\label{app:A} 

For the sake of completeness,
we provide here the approximate expressions for the density and temperature decay rates, that are relevant for
explicit computation of several of the quantities discussed in the
main text. They have been obtained from a truncated Sonine expansion
(Sonine polynomials being particular types of Laguerre polynomials,
particularly convenient for kinetic theory calculus) \cite{t02,ptd02,cdt04}.
\begin{eqnarray}
\zeta_n&=&\frac{d+2}{4}\left(1-a_2\frac{1}{16}\right),\\
\zeta_T&=&\frac{d+2}{8d}\left(1+a_2\frac{8d+11}{16}\right),
\end{eqnarray}
where
\begin{equation}\label{def:a2}
a_2=\frac{8(3-2\sqrt{2})p}{(4d+6-\sqrt{2})p+8\sqrt{2}(d-1)(1-p)}.
\end{equation}

\end{section}

\begin{section}{From the Boltzmann-Lorentz equation to the Fokker-Planck
    equation}\label{appendixA}

In this Appendix we expand the collision operator, Eq. (\ref{op_coll}), 
in series of $\Delta$. We
start by multiplying the collision operator by a generic function 
$H(\mathbf{v})$ and integrate in velocity space
\begin{eqnarray}\label{eq:a1}
&&\int\!\!d\mathbf{v}H(\mathbf{v})J[\mathbf{r},\mathbf{v},t\vert F,f]
\nonumber\\
&&=\sigma_0^{d-1}\int\!\!d\mathbf{v}\!\!\int\!\!d\mathbf{v}_1 
H(\mathbf{v})\!\!\int\!\!d\hat{\boldsymbol{\sigma}}
\Theta(\mathbf{g}\cdot\hat{\boldsymbol{\sigma}})
(\mathbf{g}\cdot\hat{\boldsymbol{\sigma}})
[F(\mathbf{v}^*)f(\mathbf{v}_1^*)-F(\mathbf{v})f(\mathbf{v}_1)].\nonumber\\
\end{eqnarray}
The above expression can be written
\begin{eqnarray}\label{eq:a2}
&&\int\!\!d\mathbf{v}H(\mathbf{v})J[\mathbf{r},\mathbf{v},t\vert F,f]
\nonumber\\
&&=\sigma_0^{d-1}\int\!\!d\mathbf{v}\!\!\int\!\!d\mathbf{v}_1
F(\mathbf{v})f(\mathbf{v}_1)\!\!\int\!\!d\hat{\boldsymbol{\sigma}}
\Theta(\mathbf{g}\cdot\hat{\boldsymbol{\sigma}})
(\mathbf{g}\cdot\hat{\boldsymbol{\sigma}})
[H(\mathbf{v}-\delta\mathbf{v})-H(\mathbf{v})],\nonumber\\
\end{eqnarray}
where we have introduced
\begin{equation}
\delta\mathbf{v}=\frac{2\Delta}{1+\Delta}
(\mathbf{g}\cdot\hat{\boldsymbol{\sigma}})\hat{\boldsymbol{\sigma}}, 
\end{equation}
which is the increment of the tagged particle velocity due to collisions
with a particle of the bath (it should be remembered that $\mathbf{g}=\mathbf{v}-\mathbf{v}_1$). 
Equation (\ref{eq:a2}) essentially tells us how
the function $H$ varies due to collisions. 
If we admit that $\Delta$ is small enough, we can expand
$H(\mathbf{v}-\delta\mathbf{v})$ around $\mathbf{v}$ in powers of
$\delta\mathbf{v}$, keeping only the lower orders
\begin{equation}\label{desarrollo1}
H(\mathbf{v}-\delta\mathbf{v})\simeq H(\mathbf{v})
-\left[\frac{\partial H(\mathbf{v})}{\partial\mathbf{v}}\right]
\cdot\delta\mathbf{v}
+\frac{1}{2}\left[\frac{\partial}{\partial\mathbf{v}}
\frac{\partial}{\partial\mathbf{v}}H(\mathbf{v})\right]
:\delta\mathbf{v}\delta\mathbf{v}.
\end{equation}
If we introduce expansion (\ref{desarrollo1}) in equation (\ref{eq:a2})
we obtain
\begin{eqnarray}\label{eq:a5}
&&\int\!\!d\mathbf{v}H(\mathbf{v})J[\mathbf{r},\mathbf{v},t\vert F,f]
\nonumber\\
&&\simeq
\int\!\!d\mathbf{v}H(\mathbf{v})\left\{\frac{\partial}{\partial\mathbf{v}}
\cdot[\mathbf{A}(\mathbf{v})F(\mathbf{v})]+\frac{1}{2}
\frac{\partial}{\partial\mathbf{v}}\frac{\partial}{\partial\mathbf{v}}
:[N(\mathbf{v})F(\mathbf{v})]\right\},
\end{eqnarray}
where we have introduced
\begin{equation}
\mathbf{A}(\mathbf{v})=\frac{2\Delta\pi^{\frac{d-1}{2}}\sigma_0^{d-1}}
{(1+\Delta)\Gamma\left(\frac{d+3}{2}\right)}
\int\!\!d\mathbf{v}_1f(\mathbf{v}_1)g\mathbf{g},
\end{equation}
\begin{equation}
N(\mathbf{v})=\left[\frac{2\Delta}{1+\Delta}\right]
\frac{\pi^{\frac{d-1}{2}}\sigma_0^{d-1}}{\Gamma\left(\frac{d+5}{2}\right)}
\int\!\!d\mathbf{v}_1f(\mathbf{v}_1)\left[\frac{d+3}{2d}g^3\mathbf{I}
+\frac{3}{2}g\left(\mathbf{g}\mathbf{g}-\frac{1}{d}g^3\mathbf{I}\right)\right].
\end{equation}
In the last expression $\mathbf{I}$ is the unit tensor. As $H(\mathbf{v})$
is a generic function of $\mathbf{v}$, we can compare equations
(\ref{eq:a1}) and (\ref{eq:a5}), and we obtain that the collision operator
can be written as
\begin{equation}\label{col_op_ex}
J[\mathbf{r},\mathbf{v},t\vert F,f]\simeq
\frac{\partial}{\partial\mathbf{v}}\cdot[\mathbf{A}(\mathbf{v})F(\mathbf{v})]
+\frac{1}{2}\frac{\partial}{\partial\mathbf{v}}
\frac{\partial}{\partial\mathbf{v}}
:[N(\mathbf{v})F(\mathbf{v})].
\end{equation}

We next specify $\mathbf{A}$ and $N$ within the scaling form 
provided by the homogeneous decay state of the bath. 
This will lead us to identify the remaining $\Delta$ dependence in 
these coefficients and to simplify the functional dependence in the 
tagged particle velocity. To this end, we introduce the
dimensionless velocities
\begin{equation}
\mathbf{c}=\frac{\mathbf{v}}{v_0(t)}, \qquad
\mathbf{c}_1=\frac{\mathbf{v}_1}{v_g(t)},
\end{equation}
where $v_g(t)=\left(\frac{2T_g(t)}{m_g}\right)^{1/2}$ and
$v_0(t)=\left(\frac{2T(t)}{m}\right)^{1/2}$, with $T(t)$ the temperature
of the tagged particle and $T_g(t)$ the temperature of the suspending gas. The relative 
velocity $\mathbf{g}=\mathbf{v}-\mathbf{v}_1$ can be written as
\begin{equation}
\mathbf{g}=v_g(t)\left[\frac{T(t)}{T_g(t)}\right]^{1/2}\Delta^{1/2}\mathbf{c}
-v_g(t)\mathbf{c}_1.
\end{equation}
A formal expansion in $(T/T_g)\Delta$ leads to
\begin{equation}\label{ec_AN}
\mathbf{A}(\mathbf{v},t)=\gamma(t)\mathbf{v}, \qquad
N(\mathbf{v},t)=2\bar{\gamma}(t)\mathbf{I}.
\end{equation}
The definitions of $\gamma$ and $\bar{\gamma}$ are respectively
\begin{eqnarray}
\gamma(t)&=&\gamma_e[n_g(t),T_g(t)]a(p),\\
\bar{\gamma}(t)&=&\gamma_e[n_g(t),T_g(t)]a(p)b(p)\frac{T_g(t)}{m},
\end{eqnarray}
where $\gamma_e(t)$ is the same friction coefficient as for an elastic system
at the corresponding density and temperature 
\begin{equation}
\gamma_e[n_g(t),T_g(t)]=\frac{4\pi^{\frac{d-1}{2}}}{d\Gamma(d/2)}\Delta^{1/2}
n_g(t)\left(\frac{2T_g(t)}{m}\right)^{1/2}\sigma_0^{d-1},
\end{equation}
and $a$, $b$ are functionals of the distribution function of the bath which
depend only on the parameter $p$
\begin{eqnarray}
a(p)&=&\frac{\Gamma(d/2)}{\Gamma((d+1)/2)}\int\!\!d\mathbf{c}_1
\chi_H(c_1)c_1,\\
b(p)&=&\frac{2}{d+1}\frac{\int\!\!d\mathbf{c}_1\chi_H(c_1)c_1^3}
{\int\!\!d\mathbf{c}_1\chi_H(c_1)c_1}.
\end{eqnarray}
\end{section}
These coefficients have been evaluated in the first Sonine approximation and depend very weakly on $p$
\begin{eqnarray}
a(p)&=&\frac{8-a_2(p)}{8},\\
b(p)&=&\frac{8+3a_2(p)}{8-a_2(p)},
\end{eqnarray}
where $a_2$, defined in (\ref{def:a2}), is the gas velocity distribution kurtosis (a Gaussian ansatz would amount to setting
$a_2=0$).
By dimensional analysis and taking into account the explicit formulas for
$\mathbf{A}$ and $N$, equation (\ref{ec_AN}), we can see that the two terms we
have considered in the expansion of the collision operator, Eq. 
(\ref{col_op_ex}), are of order $n_gv_g\sigma_0^{d-1}\Delta$, while the other
terms in the Kramers-Moyal expansion are at least of order 
$n_gv_g\sigma_0^{d-1}\Delta^{3/2}$. Hence, we can conclude that the leading
order contribution in $\Delta$ of the collision operator is actually the one
written in (\ref{col_op_ex}).

\begin{section}{Equations for the velocity and temperature of the tagged
    particle}\label{appendixB} 

In this Appendix we derive the equations for the mean velocity and temperature
of the tagged particle. Taking moments in the Fokker-Planck equation, Eq. 
(\ref{Fokker-Planck}), we obtain for the velocity
\begin{eqnarray}
\frac{\partial\mathbf{u}(t)}{\partial t}
=\int\!\!d\mathbf{r}\!\!\int\!\!d\mathbf{v}\mathbf{v}
\left\{-(\mathbf{v}\cdot\nabla)+\gamma_e(t)a(p)\frac{\partial}
{\partial\mathbf{v}}\cdot\mathbf{v}\right.\nonumber\\
\left.+a(p)b(p)\frac{T_g}{m}\gamma_e(t)\frac{\partial^2}
{\partial\mathbf{v}^2}\right\}F(\mathbf{r},\mathbf{v},t).
\end{eqnarray}
By integration we have
\begin{eqnarray}
\frac{\partial\mathbf{u}(t)}{\partial t}&=&
\int\!\!d\mathbf{r}\!\!\int\!\!d\mathbf{v}\mathbf{v}\gamma_e(t)a(p)
\frac{\partial}{\partial\mathbf{v}}
\cdot(\mathbf{v}F(\mathbf{r},\mathbf{v},t))\\
&=&-\int\!\!d\mathbf{r}\!\!\int\!\!d\mathbf{v}\gamma_e(t)a(p)\mathbf{v}
F(\mathbf{r},\mathbf{v},t)\\
&=&-\gamma_e(t)a(p)\mathbf{u}(t).
\end{eqnarray}
Consequently, the equation for the mean velocity is
\begin{equation}
\frac{\partial\mathbf{u}(t)}{\partial t}=-\gamma_e(t)a(p)\mathbf{u}(t).
\end{equation}
Taking into account the definition of temperature,
\begin{eqnarray}
\frac{d}{2}T(t)&=&\int\!\!d\mathbf{r}\!\!\int\!\!d\mathbf{v}
\frac{1}{2}m[\mathbf{v}-\mathbf{u}(t)]^2F(\mathbf{r},\mathbf{v},t)\nonumber\\
&=&\int\!\!d\mathbf{r}\!\!\int\!\!d\mathbf{v}\frac{1}{2}m[v^2-u^2(t)]
F(\mathbf{r},\mathbf{v},t).
\end{eqnarray}
we can write
\begin{equation}
\frac{d}{2}\frac{\partial}{\partial t}T(t)=\frac{m}{2}
\left[\frac{\partial}{\partial t}\int\!\!d\mathbf{r}\!\!\int\!\!d\mathbf{v}
{v}^2F(\mathbf{r},\mathbf{v},t)
-2\mathbf{u}(t)\cdot\frac{\partial \mathbf{u}(t)}{\partial t}\right].
\end{equation}
In order to evaluate the first term on the right hand side, 
we make use of the Fokker-Planck equation:
\begin{equation}
\frac{\partial}{\partial t}\int\!\!d\mathbf{r}\!\!\int\!\!d\mathbf{v}
{v}^2F(\mathbf{r},\mathbf{v},t)=-2\gamma_e(t)a(p)
\int\!\!d\mathbf{r}\!\!\int\!\!d\mathbf{v}{v}^2F(\mathbf{r},\mathbf{v},t)
+\frac{2d}{m}T_g\gamma_e(t)a(p)b(p).
\end{equation}
Taking this formula and the equation for the velocity into account, we obtain 
\begin{equation}\label{temperatura_temperaturag}
\frac{d}{2}\frac{\partial T(t)}{\partial t}=-d\gamma_e(t)a(p)[T(t)-b(p)T_g(t)].
\end{equation}

\end{section}

\begin{section}{The diffusion equation}\label{appendixC} 

In this Appendix we derive the diffusion equation for the tagged
particle's density. The starting point is  the Fokker-Planck equation (\ref{F-P-F})
\begin{equation}\label{F-P-Fa}
\frac{\partial}{\partial t^*}F_\mathbf{k}(\mathbf{v}^*,t^*)
=[\Lambda_{FP}(v^*)-i\ell_0(t^*)\mathbf{k}\cdot\mathbf{v}^*]
F_\mathbf{k}(\mathbf{v}^*,t^*),
\end{equation}
in which we introduce the two projectors
\begin{eqnarray}
Pg(\mathbf{v^*})&=&\langle\chi_M(\mathbf{v^*})\vert 
g(\mathbf{v^*})\rangle \chi_M(\mathbf{v^*}),\\
P_\perp g(\mathbf{v^*})&=&(1-P)g(\mathbf{v^*}).
\end{eqnarray}
Here, we have introduced the maxwellian distribution, $\chi_M(\mathbf{v^*}) $, which is the eigenfunction of $\Lambda_{FP}$ associated with the 0
eigenvalue and we have used the scalar product defined as
\be
\langle f(\mathbf{v}^*)\vert g(\mathbf{v}^*)\rangle=\int d\mathbf{v}^*\chi_M^{-1}(\mathbf{v}^*)f^\dagger(\mathbf{v}^*)g(\mathbf{v}^*),
\ee
$f^\dagger$ being the complex conjugate of $f$.
In a next step, we decompose the function $F_\mathbf{k}$ in $PF_\mathbf{k}$ and
$P_\perp F_\mathbf{k}$, and write the equations for these two quantities
\begin{equation}
\left[\frac{\partial}{\partial t^*}+i\ell_0(t^*)P\mathbf{k}\cdot\mathbf{v}^*
-P\Lambda_{FP}\right]PF_\mathbf{k}=-i\ell_0(t^*)P\mathbf{k}\cdot\mathbf{v}^*P_\perp F_\mathbf{k},
\end{equation}
\begin{equation}
\left[\frac{\partial}{\partial t^*}
+i\ell_0(t^*)P_\perp\mathbf{k}\cdot\mathbf{v}^*-P_\perp\Lambda_{FP}\right]
P_\perp F_\mathbf{k}=-i\ell_0(t^*)P_\perp\mathbf{k}\cdot\mathbf{v}^*P F_\mathbf{k}.
\end{equation}
We are interested in obtaining a closed equation for $PF_\mathbf{k}$ in the
hydrodynamic limit. To achieve this goal, we formally solve the equation for $P_\perp F_\mathbf{k}$ 
\begin{equation}
P_\perp
F_\mathbf{k}(\mathbf{v}^*,t^*)=G_0(t^*)F_\mathbf{k}(\mathbf{v}^*,0)
-\int_0^{t^*}dt^{*\prime}G_{t^{*\prime}}(t^*-t^{*\prime})P_\perp
i\ell_0(t^{*\prime})\mathbf{k}\cdot\mathbf{v}^* P
F_\mathbf{k}(\mathbf{v}^*,t^{*\prime}),
\end{equation}
where the operator $G_{t^{*\prime}}(t^*-t^{*\prime})$
is defined  as
\begin{equation}
\frac{d}{dt^*}G_{t^{*\prime}}(t^*-t^{*\prime})
=P_\perp[\Lambda_{FP}(\mathbf{v}^*)
-i\ell_0(t^*)\mathbf{k}\cdot\mathbf{v}^*]P_\perp 
G_{t^{*\prime}}(t^*-t^{*\prime}),
\end{equation}
with $G_{t^*}(0)=1$. In the long time limit, the term associated to the
initial condition vanishes and we have
\begin{equation}
P_\perp F_\mathbf{k}(\mathbf{v}^*,t^*)=-\int_0^{t^*} dt^*G_{t^{*\prime}}(t^*-t^{*\prime})P_\perp i
\ell_0(t^*-t^{*\prime})\mathbf{k}\cdot\mathbf{v}^* 
PF_\mathbf{k}(\mathbf{v}^*,t^*-t^{*\prime}).
\end{equation}
In order to obtain the diffusion equation to order $k^2$, we only need
$P_\perp F_\mathbf{k}$ to order $k$, so we write the expression for
$G_{t^{*\prime}}(t^*-t^{*\prime})$ to leading order
\begin{equation}
G_{t^*-t^{*\prime}}(t^{*\prime})\simeq e^{P_\perp\Lambda_{FP}P_\perp
  t^{*\prime}}. 
\end{equation}
We then have
\begin{eqnarray}
P_\perp F_\mathbf{k}(\mathbf{v}^*,t^*)&\simeq& -\int_0^{t^*}
dt^{*\prime}e^{P_\perp \Lambda_{FP}P_\perp t^{*\prime}}P_\perp i
\ell_0(t^*-t^{*\prime})\mathbf{k}\cdot\mathbf{v}^*P
F_\mathbf{k}(\mathbf{v}^*,t^*-t^{*\prime})\nonumber\\
&\simeq&- \ell_0(t^*)\int_0^{t^*} dt^{*\prime}e^{P_\perp\Lambda_{FP}P_\perp 
t^{*\prime}-\epsilon^*t^{*\prime}}P_\perp
i\mathbf{k}\cdot\mathbf{v}^* 
PF_\mathbf{k}(\mathbf{v}^*,t^*-t^{*\prime}),\nonumber\\
\end{eqnarray}
where we have used that $\ell_0(t^*)\sim e^{\epsilon^*t^{*}}$. We subsequently have to 
relate $PF_\mathbf{k}(\mathbf{v}^*,t^*-t^{*\prime})$ with $P F_\mathbf{k}(\mathbf{v}^*,t^*)$. To be consistent with the
hydrodynamic approximation, this is done to leading order 
\begin{equation}
P F_\mathbf{k}(\mathbf{v}^*,t^*-t^{*\prime})\simeq
e^{-P\Lambda_{FP}Pt^{*\prime}}PF_\mathbf{k}(\mathbf{v}^*,t^*)
=\langle\chi_M(\mathbf{v}^*)\vert
F_\mathbf{k}(\mathbf{v}^*,t^*)\rangle \chi_M(\mathbf{v}^*).
\end{equation}
Now we can write the equation for the Fourier transform of the density,
$n_\mathbf{k}(t^*)$   
\begin{equation}
\frac{\partial}{\partial t^*}n_\mathbf{k}(t^*)
=-[k\ell_0(t^*)]^2\frac{1}{d}\int d\mathbf{v}^* 
\mathbf{v}^*\cdot\int_0^{t^*} dt^{*\prime}
e^{P_\perp\Lambda_{FP}P_\perp t^{*\prime}}e^{-\epsilon^*t^{*\prime}}P_\perp
\mathbf{v}^*\chi_M(\mathbf{v}^*)n_\mathbf{k}(t^{*}).
\end{equation}
In other words,
\begin{equation}
\frac{\partial}{\partial t^*}n_\mathbf{k}(t^*)=-D_0[\ell_0(t^*)k]^2
n_\mathbf{k}(t^*). 
\end{equation}
where
\begin{equation}
D_0=\frac{1}{d}\int d\mathbf{v}^*\mathbf{v}^*\cdot\int_0^{t^*}dt^{*\prime}
e^{(P_\perp\Lambda_{FP}P_\perp-\epsilon^*)t^{*\prime}}
P_\perp\mathbf{v}^*\chi_M(\mathbf{v}^*).
\end{equation}
Finally, we can evaluate $D_0$ exactly since $v_j^*\chi_M(\mathbf{v}^*)$ is
an eigenfunction of $\Lambda_{FP}$ with eigenvalue $\lambda_1=-1$
\begin{equation}
D_0=\int d\mathbf{v}^* v_i^*\int_0^{t^*}dt^{*\prime}
e^{(-1-\epsilon^*)t^{*\prime}}P_\perp
v_i^*\chi_M(\mathbf{v}^*)=\frac{1}{2}\frac{1}{1+\epsilon^*}.
\end{equation}


\end{section}


\begin{thebibliography}{10}

\bibitem{bnklr}
E. Ben-Naim, P. Krapivsky, F. Leyvraz, and S. Redner, J. Chem. Phys. {\bf 98},
7284  (1994).

\bibitem{bek}
R. Blythe, M.~R. Evans, and Y. Kafri, Phys. Rev. Lett. {\bf 85},  3759
(2000).

\bibitem{ks}
P. Krapivsky and C. Sire, Phys. Rev. Lett. {\bf 86},  2494  (2001).

\bibitem{t02}
E. Trizac, Phys. Rev. Lett. {\bf 88},  160601  (2002).

\bibitem{ptd02}
J. Piasecki, E. Trizac, and M. Droz, Phys. Rev. E {\bf 66},  066111  (2002).

\bibitem{cdt04}
F. Coppex, M. Droz, and E. Trizac, Phys. Rev. E {\bf 70},  061102  (2004).

\bibitem{cdt04s}
F. Coppex, M. Droz, and E. Trizac, Phys. Rev. E {\bf 69},  011303  (2004).

\bibitem{llf06}
A. Lipowski, D. Lipowska, and A. Ferreira, Phys. Rev. E {\bf 73},  032102
  (2006).

\bibitem{bte05}
A. Barrat, E. Trizac, and M.~H. Ernst, J. Phys.: Condens. Matter {\bf 17},
  S2429  (2005).


\bibitem{gs95}
A. Goldshtein, and M. Shapiro, J. of Fluid. Mech. {\bf 282}, 75 (1995).

\bibitem{chapman60}
S. Chapman and T.~G. Cowling, {\em The mathematical theory of nonuniform
  gases} (Cambridge University Press, London, 1960).

\bibitem{gmsbt08}
M.~I. {Garc\'{\i}a de Soria}, P. Maynar, G. Schehr, A. Barrat, and E. Trizac 
Phys. Rev. E {\bf 77}, 051127 (2008).

\bibitem{mclennan}
J.~A. McLennan, {\em Introduction to Nonequilibrium Statistical Mechanics}
  (Prentice-Hall, Englewood Cliffs, NJ, 1989).

\bibitem{ernst}
      M.~H.~Ernst, Phys. Reports {\bf 78}, 1 (1981).

\bibitem{resibois}
P. R\'esibois and M. de~Leener, {\em Classical Kinetic Theory of Fluids} 
(John Wiley, New York, 1977).

\bibitem{TH95}
E. Trizac and J.-P. Hansen, Phys. Rev. Lett. {\bf 74},
4114 (1995). 

\bibitem{mp99}
P. A. Martin and J. Piasecki, Europhys. Lett. {\bf 46}, 613  (1999).

\bibitem{bds99}
J.~J. Brey, J.~W. Dufty, and A. Santos, J. Stat. Phys. {\bf 97}, 281  (1999).

\bibitem{brgd99}
J.~J. Brey, M. J. Ruiz-Montero, R. Garcia-Rojo and J.~W. Dufty, 
Phys. Rev. E {\bf 60},  7174  (1999).

\bibitem{dbl02th}
 J.~W. Dufty, J.~J. Brey, and J. Lutsko
Phys. Rev. E {\bf 65}, 051303 (2002).

\bibitem{bt02}
A, Barrat and E. Trizac,
Granular Matter  {\bf 4}, 57 (2002).

\bibitem{sd06}
A. Santos, and J. W. Dufty, Phys. Rev. Lett. {\bf 97}, 058001  (2006).

\bibitem{dg01}
J.~W. Dufty, and V. Garz\'o, J. Stat. Phys. {\bf 105}, 723  (2001).

\bibitem{blp04}
A. Barrat, and V. Loreto, and A. Puglisi, Physica A {\bf 334}, 513 (2004). 

\bibitem{g04}
V. Garz\'o, Physica A {\bf 343}, 105 (2004). 

\bibitem{bird}
G.~A. Bird, {\em Molecular Gas Dynamics and the Direct Simulation of Gas Flows}
  (Clarendon, Oxford, 1994).



\end{thebibliography}
\end{document}